\documentclass[sigconf, nonacm]{acmart}

\usepackage{graphicx}
\usepackage{balance}
\usepackage{algorithm}
\usepackage{algpseudocode}
\usepackage{fix-cm}
\usepackage{multirow}
\usepackage{caption}
\usepackage{subcaption}
\usepackage{amsmath}
\usepackage{bbold}
\usepackage{mathtools}
\usepackage{xcolor}
\usepackage{threeparttable}
\usepackage{hyperref}
\usepackage{amsthm}
\usepackage{enumitem}
\usepackage{pifont}
\usepackage{gensymb}
\usepackage{listings}
\usepackage{subfiles}
\usepackage{xr} 
\usepackage{wrapfig}
\usepackage{setspace}
\usepackage{booktabs}
\usepackage[export]{adjustbox}
\usepackage{xspace}
\usepackage{footnote}
\makesavenoteenv{tabular}
\makesavenoteenv{table}

\usepackage[colorinlistoftodos,prependcaption,textsize=tiny]{todonotes}
\usepackage{varwidth}
\newcommand{\META}[1]{\ignorespaces}
\newcommand{\RI}[1]{\ignorespaces}
\newcommand{\RII}[1]{\ignorespaces}
\newcommand{\RIICAP}[1]{\ignorespaces}
\newcommand{\RIII}[1]{\ignorespaces}
\newcommand{\RIIICAP}[1]{\ignorespaces}

\newtheorem*{definition}{DEFINITION}

\algnewcommand\algorithmicforeach{\textbf{for each}}
\algdef{S}[FOR]{ForEach}[1]{\algorithmicforeach\ #1\ \algorithmicdo}
\algnewcommand\And{\textbf{and }}
\algnewcommand\Or{\textbf{or }}

\usepackage{xcolor}  %
\definecolor{Plum}{HTML}{C2938D}
\definecolor{light-gray}{gray}{0.8}
\newcommand{\deleted}[1]{{\color{light-gray}#1}}
\newcommand{\added}[1]{#1}

\renewcommand{\deleted}[1]{}

\newcommand{\xindex}{XIndex\xspace}
\newcommand{\pgm}{PGM-Index\xspace}
\newcommand{\fiting}{FITing-Tree\xspace}
\newcommand{\finedex}{FINEdex\xspace}
\newcommand{\alex}{ALEX\xspace}
\newcommand{\alexplus}{ALEX+\xspace}
\newcommand{\apex}{APEX\xspace}
\newcommand{\lipp}{LIPP\xspace}
\newcommand{\lippplus}{LIPP+\xspace}
\newcommand{\art}{ART\xspace}
\newcommand{\hot}{HOT\xspace}

\definecolor{deepblue}{rgb}{0,0,0.5}
\definecolor{deepred}{rgb}{0.6,0,0}
\definecolor{deepgreen}{rgb}{0,0.5,0}

\lstdefinestyle{mystyle}{
    commentstyle=\color{codegreen},
    keywordstyle=\bf\color{deepblue},
    numberstyle=\tiny\color{black},
    stringstyle=\color{deepgreen},
    basicstyle=\ttfamily\footnotesize,
    breakatwhitespace=false,         
    breaklines=true,                 
    captionpos=b,                    
    keepspaces=true,                 
    showspaces=false,                
    showstringspaces=false,
    showtabs=false,                  
    tabsize=2
}
\usepackage{array}
\newcolumntype{M}[1]{>{\centering\arraybackslash}m{#1}}
\newcolumntype{P}[1]{>{\centering\arraybackslash}p{#1}}

\newcounter{msg}
\newenvironment{msg}{\refstepcounter{msg}\par\addvspace{\smallskipamount}
   \noindent \textbf{Message \themsg.} \bfseries}{\par\addvspace{\smallskipamount}}

\begin{document}
\setlength\emergencystretch{\hsize}

\title{Are Updatable Learned Indexes Ready?
}

\author{Chaichon Wongkham}
\affiliation{\institution{The Chinese University of Hong Kong} \country{}}
\email{wongkha21@cse.cuhk.edu.hk}
\author{Baotong Lu}
\affiliation{\institution{The Chinese University of Hong Kong} \country{}}
\email{btlu@cse.cuhk.edu.hk}
\author{Chris Liu}
\affiliation{\institution{The Chinese University of Hong Kong} \country{}}
\email{cyliu@cse.cuhk.edu.hk}
\author{Zhicong Zhong}
\affiliation{\institution{The Chinese University of Hong Kong} \country{}}
\email{zczhong@cse.cuhk.edu.hk}
\author{Eric Lo}
\affiliation{\institution{The Chinese University of Hong Kong} \country{}}
\email{ericlo@cse.cuhk.edu.hk}
\author{Tianzheng Wang}
\affiliation{\institution{Simon Fraser University} \country{}}
\email{tzwang@sfu.ca}

\begin{abstract}
Recently, numerous promising results have shown
that updatable learned indexes can perform better than traditional indexes with much lower memory space consumption.  
But it is unknown how these learned indexes compare against each other and against the traditional ones under realistic workloads with changing data distributions and concurrency levels. 
This makes practitioners still wary about how these new indexes would actually behave in practice. 
To fill this gap, %
this paper conducts the first comprehensive evaluation on updatable learned indexes. 
Our evaluation uses ten real datasets and various workloads to challenge learned indexes in three aspects: performance, memory space efficiency and robustness. 
Based on the results, 
we give a series of takeaways that can guide the future development and deployment of learned indexes.

\end{abstract}

\maketitle

\section{Introduction} 
\label{sec:introduction}
Recent advances in machine learning (ML) have 
sparked a flurry of research  on
using ML to improve various database components. 
However, it remains unclear (1) what the relative merits are of the various techniques and (2) how those impressive 
results will hold up under various workloads. %
Practitioners are often wary about using these new techniques in their systems %
without knowing how they would perform in practice.
In this paper, we hope to fill the gap 
by studying \emph{learned indexes}~\cite{rmi, alex, apex, radix, fiting, lipp, xindex, pgm, bourbon,sindex,shifttable,lisa, tsunami,sprig,RSMI,qd-tree, ZM,ML-Index, md-li, book},
the arguably most well-studied learned database engine component. 

Although preliminary evaluations on learned indexes exist, they largely focused on read-only workloads~\cite{BenchmarkingLearnedIndexes-vldb,JensRMI}.
A system built for practical adoption, however, must also consider dynamic workloads.
Hence, we focus on \emph{updatable} learned indexes.
Moreover, to provide a holistic evaluation, we focus not only on (1) performance (throughput and latency), but also (2) space efficiency and (3) robustness under dynamic workloads. 
Next, we briefly introduce our evaluation methodology and highlight our findings.

\subsection{Performance} 
\label{subsec:perf}
Since learned indexes 
can easily 
overfit to a particular synthetic distribution \cite{TowardsBenchmarkforLearnedSystems},
the SOSD benchmark \cite{sosd} has taken the initiative with four real datasets for evaluating learned indexes. 
Recent evaluations have started using SOSD's real datasets \cite{BenchmarkingLearnedIndexes-vldb,JensRMI}, 
but they only focused on read-only workloads.
Among the four real datasets in SOSD,
{\tt osm} and {\tt facebook} are known to be more challenging for learned indexes~\cite{JensRMI,BenchmarkingLearnedIndexes-vldb}.
In contrast, learned indexes are able to outperform traditional indexes on
the {\tt wiki} and {\tt books} datasets~\cite{JensRMI,BenchmarkingLearnedIndexes-vldb}.
However, it is unclear how these results (i.e., which type of index performs better under which scenarios) may extend beyond the four concrete datasets, leading to the question: \textit{can a learned index still outperform all traditional indexes on a variety of data?}
We observe the fundamental reason is the lack of a common, \textit{quantitative} metric to evaluate how ``hard'' (or ``easy'') a dataset is for a learned index to handle.
Furthermore, without a quantitative measure, it is hard for practitioners to know whether 
their target workload/production data is ``easy'' or ``hard'', which may eventually impact which index should be chosen to achieve the best performance.

In this paper, we propose to use \emph{piecewise linear approximation} (PLA)~\cite{pgm-geometry-paper}
as an approximate metric to quantify the hardness of a dataset.
PLA has a clear theoretical foundation from computational geometry,
which captures the minimal number of linear models required to fit a data distribution.
It has been used to build a space-optimal learned index~\cite{pgm}. %
We use PLA to approximate the hardness of a dataset 
in two dimensions: \emph{global hardness} and \emph{local hardness}. 
The former challenges the index's structure design choices and the cost models that govern structural modification operations (SMOs); the latter challenges the accuracy of various ML models.

The global and local hardness metrics based on PLA enable us to conduct a comprehensive study 
that pragmatically examines 
updatable learned indexes along three aspects:
(1) {\bf data}  (from \emph{easy} to \emph{hard}),
(2) {\bf workload} (from \emph{read-only} to \emph{write-only}), and
(3) {\bf concurrency} (from \emph{single-threaded/single-socket} to \emph{multi-threaded/NUMA}).
Our results revealed a number of interesting insights.
For example, although our results show that 
recent updatable learned indexes (ALEX \cite{alex}, LIPP \cite{lipp}) 
perform better 
than traditional indexes (ART \cite{art}, Masstree \cite{masstree}, HOT \cite{hot} and Wormhole \cite{wormhole}) in over 80\% of the data-workload space under a single thread, 
some learned index design choices are actually at odd with 
concurrency control and show regression under multiple threads.

\subsection{Space Efficiency}
Memory consumption is an important factor in production environments.
Learned indexes have been reported to use 4$\times$-2000$\times$ less memory compared to  traditional indexes because 
they store succinct models instead of keys in nodes~\cite{rmi,alex}.
While their space advantage has been proven in read-only environments,
we found that it might not hold up in a realistic environment with data modifications.
In read-only workloads,
the data array to be indexed can be fully ordered and packed, so 
even the leaf nodes can store no keys: they can store and 
use a model to point straight to the underlying data array.
In contrast, a real system 
often needs to handle dynamic workloads and builds indexes on non-primary keys.
In both cases, the underlying data array could be unsorted or the nodes have to leave spaces to accommodate insertions.
Consequently, the index often
needs a leaf layer that explicitly stores the key-position information, bloating memory usage.

In this paper, we examine the \emph{end-to-end} memory space efficiency 
of learned indexes 
including their leaf layer with key-position pairs.
We found that learned indexes are only 3.2$\times$ smaller than
traditional indexes at best and might even use more space than some traditional indexes.
Note that traditional indexes can consume around 55\% of the total memory in transactional in-memory databases \cite{HybridIndex}.
Therefore, our results suggest that memory saving 
is not a definitive advantage of the current generation of updatable learned indexes and highlight the need for further improvements in this area. 

\subsection{Robustness}

Modern data systems often 
require predictable performance. 
Low tail latency is particularly important for 
user-facing applications~\cite{lowtail}.
In this paper, we examine whether the high performance of learned indexes 
comes with a price on poor tail latency.
Furthermore, %
many workloads require efficient and robust support for range queries. 
In addition to point queries, we therefore also test learned indexes' behavior under varying range scan sizes and compare them with %
state-of-the-art traditional indexes. %
Finally, the power of learned indexes 
is that they allow specialization to a given data distribution,
hence becoming \emph{instance-optimized}~\cite{TowardsBenchmarkforLearnedSystems}.
However, in a realistic environment, the data can continuously evolve. 
It then becomes important to study how well and how fast a learned index can adapt to changes.
Yet existing performance studies~\cite{JensRMI,sosd,BenchmarkingLearnedIndexes-vldb} have
not looked into this issue.
Some recent updatable learned indexes, such as ALEX~\cite{alex} 
and XIndex~\cite{xindex}, have briefly touched upon this issue but did so without comparing with a wide range of traditional/learned indexes as well as on different data hardness.
In this paper, we study the impact
of data distribution changes on 
learned indexes and compare their robustness with that of the traditional indexes.
Our results show that although a few learned indexes generally showed good resilience to changes in data distribution, most traditional indexes exhibit even better, extremely robust performance across different workloads and data distributions.

\subsection{Contributions, Limitations, and Roadmap} 

We have created and open-sourced a benchmarking suite, GRE\footnote{Akin to the GRE test %
for a learned index to ``pass and graduate'' for practical use. Available at \url{https://github.com/gre4index/GRE}.}
which makes it easy to compare against learned and traditional indexes
under the evaluation spectrum proposed in this paper. 
To facilitate future studies,
GRE includes
scripts to run the benchmark and visualize all experiments presented in this paper.
While earlier effort~\cite{BenchmarkingLearnedIndexes-vldb} 
has presented a leaderboard for learned indexes,
GRE further expands the dimension of the comparisons by presenting \emph{heatmaps}
to visualize which ``area'' in the data-workload spectrum
has already been ``conquered'' by learned indexes and which area the traditional indexes are still in possession.

To the best of our knowledge, 
this is the first comprehensive study of updatable learned indexes' performance, space efficiency and robustness. 
However, since learned indexes are still fast evolving with new proposals on handling multi-dimensional data \cite{tsunami, md-li,lisa,RSMI,qd-tree, ZM,ML-Index}, persistence~\cite{apex} and string keys~\cite{sindex, umar21}, 
we focus on \added{one-dimensional updatable indexes on numeric data} 
and anticipate future work to cover the other aspects.

In the rest of this paper, 
Section \ref{sec:background_related_work} provides the background on %
updatable learned indexes.  
Section \ref{sec:exp} presents our experimental setup, %
including how we approximate the hardness of a dataset.
Sections~\ref{sec:perf}--\ref{sec:syn} then present the detailed empirical results. 
Section~\ref{sec:lessons} summarizes our lessons learned
and then we answer the theme question \textit{Are updatable learned indexes ready?} in Section \ref{sec:conclusion}.
Due to space limit, more discussions and results could be found in \cite{gre}.

\section{Learned Indexes}
\label{sec:background_related_work} 
The main intuition behind learned indexes 
is that if 
the data is a set of continuous integer keys (e.g., the keys 0 to 100M) in an array~\texttt{a}, 
the key itself can be directly used as an offset into the array (e.g., the value of key 77 can be accessed at \texttt{a[77]}). 
This can potentially allow O(1) rather than O(log n) lookup complexity and significantly reduce index storage overhead. 
However, implementing a practically useful learned index is challenging in terms of supporting non-contiguous data, dynamic workloads, and concurrency. 

\subsection{Structure and Last-Mile Search}
The ideal case with O(1) complexity is based on a crucial assumption that keys are continuous, which is not necessarily true in reality. 
Therefore, learned indexes usually train multiple ML models to approximate the 
cumulative distribution function (CDF) of the data, forming a hierarchy of models akin to the structure of B+-trees with inner and leaf nodes.
However, instead of laying out keys physically, inner nodes in learned indexes store models. 
The model guides traversals to reach leaf nodes which store the actual data. %
If simple linear models are used, the node only needs to store the slope and intercept, reducing storage footprint. \RI{R1: W3/D3/D7/I3}
\added{There are two inner node designs for updatable learned indexes:

\textbf{ML for Child Search.}
In this design, an inner node $I$
uses a model to predict which child node of $I$ contains the key.
Hence, the model of $I$ is trained using the information from its children (e.g., maximum and minimum keys of each child).
Since a model might be imperfect and give wrong predictions,
it is necessary to search around the predicted position to reach the exact child.
PGM-Index~\cite{pgm}, XIndex~\cite{xindex} and FINEdex~\cite{finedex} adopt this  design.

\textbf{ML for Subspace Lookup.}  
Indexes using this design (ALEX~\cite{alex}, APEX~\cite{apex} and LIPP~\cite{lipp}) recursively partition the key space, 
and a partition is represented by a slot in an inner node.
A model in each inner node is used to \emph{decide} which partition a key belongs to.
Thus, a traversal simply uses the model
to \emph{compute} which child (partition) shall be visited next,
instead of predicting.
}

When reaching a leaf node,
there is always a ``last-mile'' search around the model predicated position 
because models may not be perfect.
Last-mile search often is a performance bottleneck. 
Different learned indexes use different designs to mitigate it:

\textbf{Error-Driven Designs. } 
This type of updatable learned indexes requires as input an error threshold to bound the distance of the last-mile search; 
it is adopted by PGM-Index~\cite{pgm}, \xindex~\cite{xindex}, FITing-Tree~\cite{fiting} and FINEdex~\cite{finedex}.
In case a model's error margin goes beyond the given threshold, the index will carry out adjustment such as increasing the granularity of the models~\cite{radix} or increasing the number of nodes (and hence increasing the number of models since each node hosts one model)~\cite{pgm,xindex,fiting}.

\textbf{Performance-Driven Designs.} 
ALEX~\cite{alex} and APEX~\cite{apex} maintain runtime statistics (e.g., the number of keys searched in the last-mile and tree height) to detect suboptimal behaviors due to model inaccuracies.
These indexes maximize the accuracy by triggering the best action (e.g., SMOs) based on an empirical-based cost model.

\textbf{Collision-Driven Designs.} 
The key idea is to model inaccuracies as ``collisions.'' 
Specifically, a model-based CDF can be seen as an order-preserving function where different keys could be ``hashed'' (predicted) to the same ``bucket'' (array position), causing collisions.
Hence, a last-mile search could be handled by an open-addressing collision resolution scheme that gets or puts a collided key elsewhere in the index. 
Based on this idea, LIPP~\cite{lipp} attempts to eliminate last-mile search by finding a model that minimizes collisions. %
When a collision is inevitable, it uses 
a chaining scheme that creates new 
nodes (or recursively a sub-tree of new nodes) to transform the last-mile search problem to be a sub-tree traversal problem.

\subsection{Dynamic Workloads}
\label{subsec:dynamic}
Beyond lookups, updatable learned indexes also need to support inserts/deletes/updates which can change data distributions. 
Existing solutions usually use tree-merge, delta-merge or sparse nodes. 

\textbf{Tree-Merge.}
As represented by \pgm~\cite{pgm}, tree-merged based approaches are inspired by LSM-trees~\cite{lsm} which build multiple sub-indexes, each of which covers a subset of keys.
Inserts are then handled by creating new sub-indexes by merging smaller sub-indexes with the new key to ensure the keys stay sorted. Updates are done in-place and deletes are implemented as inserts of tombstones.
Consequently, tree-merge based designs inherit the performance characteristics of LSM-trees, for example, searching for a key may visit multiple sub-indexes of different sizes. 

\textbf{Delta-Merge.} Unlike tree-merge designs, in \xindex~\cite{xindex}, \fiting~\cite{fiting} and \finedex~\cite{finedex}, %
new keys are collected in delta nodes and are merged periodically with the main data array with potential model retraining. 
Updates and deletes are done in-place.
The main difference among the various indexes under this design
is their delta granularity.
For \xindex, new inserts are absorbed by a per-node delta.
\finedex~\cite{finedex} maintains a delta per record to reduce conflicts within a node and facilitate parallel retraining.

\textbf{Sparse-Nodes.} To absorb future inserts, \alex~\cite{alex}, \apex~\cite{apex} and \lipp~\cite{lipp} leave space in tree nodes, making them sparse.  
Updates and deletes are also done in-place.
The index then needs to conduct SMOs 
such as node merges and splits, similar to traditional indexes. 
Moreover, to balance between the tree's fill factor and model accuracy, the %
index should
consider learned-index specific operations, such as node resizing and model retraining when the workload and data distribution change.

\subsection{Concurrency}
\label{subsec:concurrency}
Since most learned indexes employ a hierarchy of models, classic concurrency control approaches for trees can be adapted to work on learned indexes.
Among the surveyed in-memory learned indexes, only 
\finedex~\cite{finedex} and \xindex~\cite{xindex} support concurrency. %
To cope with the in-memory environment, 
both of them use
optimistic locking, 
which associates a versioned lock per node. 
The lock word carries a version number, such that readers only need to verify the version did not change before and after their accesses; 
writers would acquire locks as usual 
but also increment the version number. 
This extracts more concurrency by allowing readers to proceed (especially, to traverse inner nodes) without holding locks. %

The use of optimistic concurrency has impact on memory reclamation and delta merging procedures. 
For example, in \xindex, before merging a node's delta with %
its main data array,
it uses 
RCU~\cite{epoch} to  
ensure the existing readers of the delta have all finished.
During the merge, writers follow the normal write path but inserts would work on yet another temporary per-node delta (so readers also need to read this temporary delta in additional to the index before merge; \xindex's current implementation uses Masstree~\cite{masstree} to construct the delta). 
Finally, the temporary delta would be promoted as the official delta after the merge is done.
Using this approach, both readers and writers are non-blocking.

\section{Benchmarking Setup} \label{sec:exp}

We conduct experiments using various datasets and workloads, and compare them against state-of-the-art traditional indexes. 
To set the stage, we describe the necessary changes done to each index, datasets and workloads, and our environment.

\begin{table}[t]
\caption{\added{Configurations of learned indexes. Indexes named with a ``+'' sign are our concurrent implementations.}}
\centering
\scalebox{0.8}{
\begin{tabular}{ ll } 
\hline
\textbf{Index} & \textbf{Parameters} \\\hline
ALEX & Max inner/data node size: 16MB \\
& Min/avg/max node density: 0.6/0.7/0.8 \\ 
\hline
ALEX+ & Max inner/data node size: 16MB/512KB \\
&  Min/avg/max node density: 0.6/0.7/0.8 \\ 
\hline
LIPP(+) & Node density: 0.5; max node size: 16MB \\
& Subtree inserted/conflict ratio: 2/0.1 \\
\hline
PGM-Index\footnote{PGM-index cannot support key in value $2^{64}-1$.  {\tt fb} contains that key. Following \url{https://github.com/gvinciguerra/PGM-index/issues/29}, we shifted all {\tt fb} keys by -1 when using PGM-Index.
Furthermore, the codebase of PGM-Index has continuously evolved since its publication.  The latest version exposes 3 extra knobs. We tuned PGM-Index based on the authors' recommendation.} & Error bound: 16 \\
\hline
XIndex & Error bound: 32; delta size: 256 \\
&Error tolerance: 1/4; max number of models per group: 4 \\
\hline
FINEdex & Error bound: 32 \\\hline
\end{tabular}
}
\label{tbl:index_config}
\end{table}

\subsection{Index Implementations}
Since the implementation of a learned index can highly influence experimental results~\cite{sosd},
we follow previous work~\cite{BenchmarkingLearnedIndexes-vldb,JensRMI,ModernIndexICDE,PMRangeIndex} to use the original authors' or widely-used open-source implementations.
\RII{AE7 R2: W3/D5/I3}
\added{Unless otherwise stated, we use the recommended parameters that can be found in each index's original implementation. 
All the evaluated learned indexes along with their configurations are listed in Table~\ref{tbl:index_config}.}
Among the included learned indexes, 
\lipp and \alex do not support concurrency out-of-the-box, but our evaluation shows that they are the most competitive under a single thread. 
So we implemented concurrent versions of them (\lippplus and \alexplus) to make  comprehensive comparisons and reason about their design decisions in realistic multi-core environments.

\lipp recommended lock coupling~\cite{lc} for concurrency control~\cite{lipp}.
However, coupling using traditional locks may severely limit concurrency~\cite{olc}. 
Although using optimistic lock coupling~\cite{olc} can mitigate the impact, we observe that \lipp actually requires no coupling because when following a pointer from a parent node to a child node, the pointer would not be invalidated by another concurrent thread.
Yet, LIPP does not differentiate inner node and leaf node
but uses a unified node layout: a node can store both 
data and pointers to child nodes. 
Unfortunately, this design choice limits scalability
because 
\added{(i) inner nodes now contain keys and must also maintain statistics for SMO decision making (where ALEX only needs to maintain statistics for leaf nodes)}
and 
(ii) even a simple key insert that triggers no SMOs could  
lock a node at any level (e.g., the root).
We mitigate the latter by using item-level optimistic locks (without coupling) for \lipp (denoted as \lippplus), where reads proceed without taking the lock but only need to verify that the read item (i.e. data or child pointer) did not change; 
only writers are required to take the lock in exclusive mode. 
Furthermore, we uncovered two bugs in LIPP's original implementation, fixing which 
led to slightly lower performance compared to what was originally reported~\cite{lipp}.\footnote{
Our fixes have been accepted by the original authors (details at \url{https://github.com/Jiacheng-WU/lipp/pull/11} and \url{https://github.com/Jiacheng-WU/lipp/pull/12}).}

The original implementation of \alex did not support concurrency. 
We implemented a concurrent version (\alexplus) by adapting the concurrency protocol of \alex's persistent memory variant, \apex~\cite{apex}. 
\apex employs optimistic locks at leaf node level for every 256 records.
By employing out-of-place-based SMOs (i.e., always allocating new nodes), \apex traverses tree without holding locks. %
Inner nodes are synchronized using per-node shared-exclusive locks.
\alexplus adopts all the concurrency design from APEX except 
it uses a single optimistic lock per data node,
which exhibits better performance in DRAM environments. 
We excluded FITing-tree~\cite{fiting} in our experiments 
since it is not open-source.

For traditional indexes, we include STX B+-tree~\cite{btreestx}, ART \cite{art} and HOT \cite{hot} for single-threaded experiments. 
For multi-threaded experiments, we include B+-TreeOLC~\cite{btreeolc}, ART-OLC (ART with optimistic lock coupling)~\cite{olc}, HOT-ROWEX (HOT with ROWEX~\cite{olc}), Masstree \cite{masstree} and Wormhole \cite{wormhole}.
Due to space limitation, we omit their details and only cover the necessary changes made by us. 
We added side-links in leaf nodes for B+-TreeOLC for better range scan performance. 
Two ART implementations support concurrency; we use the best performing one based on OLC.  
We ported HOT's epoch-based memory reclamation (EBMR) to ART for better scalability.
Although HIST-tree~\cite{histtree} shows that as a traditional index, it can achieve promising speedups by leveraging 
certain implicit assumptions made by learned indexes (e.g., sortedness), its open-source implementation does not support dynamic workloads and we exclude it from our study. 

\subsection{Datasets}
Table~\ref{tbl:data} shows the real datasets used in our benchmarks. 
The first four datasets are from SOSD~\cite{sosd}.
{\tt osm} from SOSD is known to be a
one-dimensional projection of multi-dimensional spatial data.
We include it for stress testing under the worst case scenarios.
Except {\tt wiki}, each dataset consists of 200M unique 8-byte unsigned integer keys, and we pair each key with an 8-byte payload. 
{\tt wiki} has duplicated keys.\RII{AE6 R2: W1/D3/I2}

Some of these datasets have been used by prior studies but it remains hard to quantify their potential impact on index performance. 
For example, some studies have shown that learned indexes cannot outperform traditional indexes on the {\tt osm} dataset~\cite{BenchmarkingLearnedIndexes-vldb}. 
Other results have shown that learned indexes can outperform traditional indexes under {\tt wiki} and {\tt books} in read-only workloads~\cite{BenchmarkingLearnedIndexes-vldb,JensRMI}.
This leads to questions such as \textit{how ``hard'' is {\tt osm} actually?} and \textit{is it much harder or just a bit harder than {\tt books}?}
We thus observe the need to quantify the ``hardness'' of a dataset. %
As we detail next, we find that piecewise linear approximation (PLA)~\cite{pgm} is a good start. %

\begin{table}[t]
\caption{Datasets used in experiments.}
\label{tbl:data}
\scalebox{0.8}{
\begin{tabular}{llc}\hline
\bf Dataset &  \bf Description  & \bf Source \\ \hline
{\tt books} & Amazon book sales popularity  &  \cite{sosd} \\
{\tt fb} & Upsampled Facebook user ID  & \cite{sosd} \\
{\tt osm} &  Uniformly sampled OpenStreetMap locations &  \cite{sosd} \\
\added{{\tt wiki}} & \added{Wikipedia article edit timestamps} &  \cite{sosd} \\
{\tt covid} & Uniformly sampled Tweet ID with tag COVID-19 &  \cite{covid} \\
{\tt genome} & Loci pairs in human chromosomes  & \cite{genome} \\
{\tt stack} & Vote ID from Stackoverflow  & \cite{stack} \\
{\tt wise} & Partition key from the WISE data & \cite{wise} \\
{\tt libio} & Repository ID from libraries.io &  \cite{libio} \\
{\tt history} & History node ID in OpenStreetMap  &  \cite{osm} \\
{\tt planet} & Planet ID in OpenStreetMap & \cite{osm} \\\hline
\end{tabular}
}
\end{table}

All 1D learned indexes regard indexing as mappings from keys to positions. 
If the data distribution is more non-linear, 
learned indexes will use more linear models to approximate the distribution.
Hence, we could use the minimum number of linear models required to fit the distribution to approximate the data hardness: 

\begin{definition}[$\epsilon$-approximate \cite{pgm}]
Given an array $D = [k_1, k_2, \\ \ldots, k_n]$,
where $k_i$ is a key with rank $r_i$ in the array,
if a model $F$ for array $D$ is $\epsilon$-approximate, 
then $|F(k_i)-r_i| \le \epsilon$, $\forall i \in [1,n]$.
\end{definition}

A given array $D$ may not be perfectly fitted by a single $\epsilon$-approximate linear model.  %
But one could split $D$ into multiple small segments $D_1,D_2, \ldots, D_m$ such that for each $D_i$ there exists an $\epsilon$-approximate linear model $F_i, i\in[1,m]$.
The minimal collection of those models is then the PLA of $D$.
\added{Computing the optimal PLA model of $D$ can be done in linear time \RI{R1: W2/D2/I2}
using the algorithm in \cite{pgm}.} %
For a given array $D$ and an error bound $\epsilon$, 
we define the data hardness $H$ as the number of segments 
in $D$'s optimal PLA model, which includes the least number of segments.
For the same dataset, $H$
should increase as $\epsilon$ decreases. 
In the rest of this paper, we use $\epsilon=4096$ and  $\epsilon=32$ 
to quantify the hardness of a dataset in two dimensions.

\begin{figure}
    \centering
     \begin{subfigure}{0.45\linewidth}
         \centering
         \includegraphics[width=\linewidth]{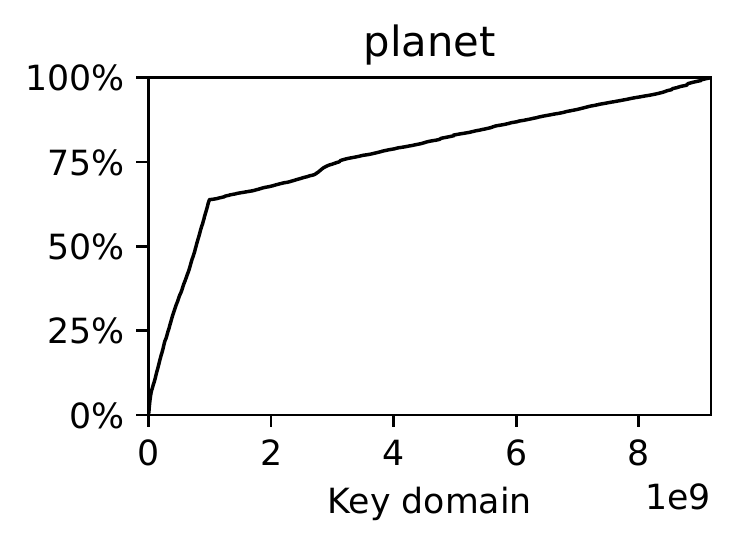}
     \end{subfigure}
     \begin{subfigure}{0.45\linewidth}
         \centering
         \includegraphics[width=\linewidth]{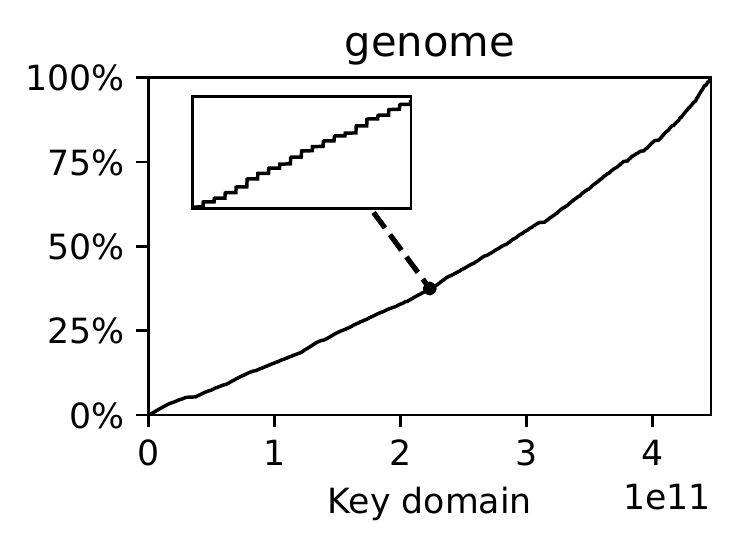}
    \end{subfigure}
    \caption{CDFs of {\tt planet} and {\tt genome}}
    \label{fig:data}
\end{figure}

When $\epsilon$ is large (4096), the PLA is more coarse-grained and 
can capture the dataset's global non-linearity, 
which mainly 
challenges the structural aspect of a learned index. 
Specifically, 
since learned indexes conceptually break down a CDF as a tree of models, 
its structure (e.g., fanout and height) is strongly influenced by the global non-linearity of the CDF.
For example, our experimental results (Section \ref{sec:exp}) indicate that many learned indexes cannot outperform the traditional ones 
on the {\tt planet} dataset.
By looking at its CDF (Figure \ref{fig:data}a), 
we see that its hardness comes from 
its sharp deflection of distribution at key value around 1M,
giving it high global non-linearity.
On datasets like this,
learned indexes that insist on a balanced tree structure (e.g., PGM-Index) 
would give the dense key region (keys $<$1M) and 
the sparse  key region (keys $>$1M) the same height,
requiring more layers to be traversed in the sparse key region than needed.
ALEX and LIPP combat the issue by adopting unbalanced trees (hence the dense region, if hard-to-fit, can be approximated with more models).
Nonetheless, a CDF like the one in Figure \ref{fig:data}a is still challenging %
since it can increase their path length variance (affecting latency)
and stress their decision components (e.g., the cost model in ALEX).

Global non-linearity alone is insufficient to fully characterize 
the hardness of a dataset to learned indexes. 
For example, we found that the {\tt genome} dataset 
also gives a hard time to many learned indexes.
However, if looking at its CDF (Figure \ref{fig:data}b), its CDF looks smooth.
But if we zoom into its CDF,
it is found that {\tt genome} has a very bumpy distribution locally.
Non-linearities in local regions 
challenge the individual machine learning models of a learned index.
Hence, we use PLA with a small $\epsilon$ (32) 
to capture local non-linearity.

Global non-linearity (PLA $\epsilon$-4096) and 
local non-linearity (PLA $\epsilon$-32) together form our \emph{data space} 
that quantifies the hardness of a dataset.
The $\epsilon$ values 4096 and 32 are empirically decided.
We have also tried other metrics to approximate a dataset's hardness, 
e.g., by measuring the mean square error of fitting only one linear regression line.
But %
we found that PLA $\epsilon$-4096 and PLA $\epsilon$-32 aligned the best with the actual index performance 
when there are no other factors (e.g., NUMA, concurrency control) in play.

\subsection{Workloads}
We devise synthetic workloads that issue requests using the aforementioned datasets. 
For each dataset, we first randomly shuffle all the 200 million keys, and then issue lookup/insert requests according the specified ratios below. 

\added{\RIII{R3 D2}
\begin{itemize}[leftmargin=*]\setlength\itemsep{0em}
\item \texttt{Read-Only (0\% Write)}: Bulk load all the 200M keys and randomly lookup for 800M keys.  %
\item \texttt{Read-Intensive (20\% Write)}: Bulk load 100M random keys, then issue requests 
where 80\% are lookup operations and 20\% are insert operations that insert all the remaining keys. %
\item \texttt{Balanced (50\% Write)}: Same as \texttt{Read-Intensive} but 50\% are lookup operations and 50\% are insert operations.
\item \texttt{Write-Heavy (80\% Write)}: Same as \texttt{Balanced} but the lookup/insert ratio is 20\%:80\%.
\item \texttt{Write-Only (100\% Write)}: Issue 100 million insertion after bulk loading 100 million keys. %
\end{itemize}
}

For each workload and index,
we repeat the experiment three times 
and report the average throughput (operations per second) and the average latency after bulk loading.

\subsection{Hardware and Platform}
All the experiments are conducted on a quad-socket machine with four 24-core Intel Xeon Platinum 8268 CPUs clocked at 2.9 GHz.
The machine in total has 96 cores 
and 768GB of main memory. 
By default, hyper-threading is disabled.
All the code is compiled using gcc 8.3.0 under the \texttt{O3} optimization level.

\begin{figure}
    \centering
    \includegraphics[width=0.88\linewidth]{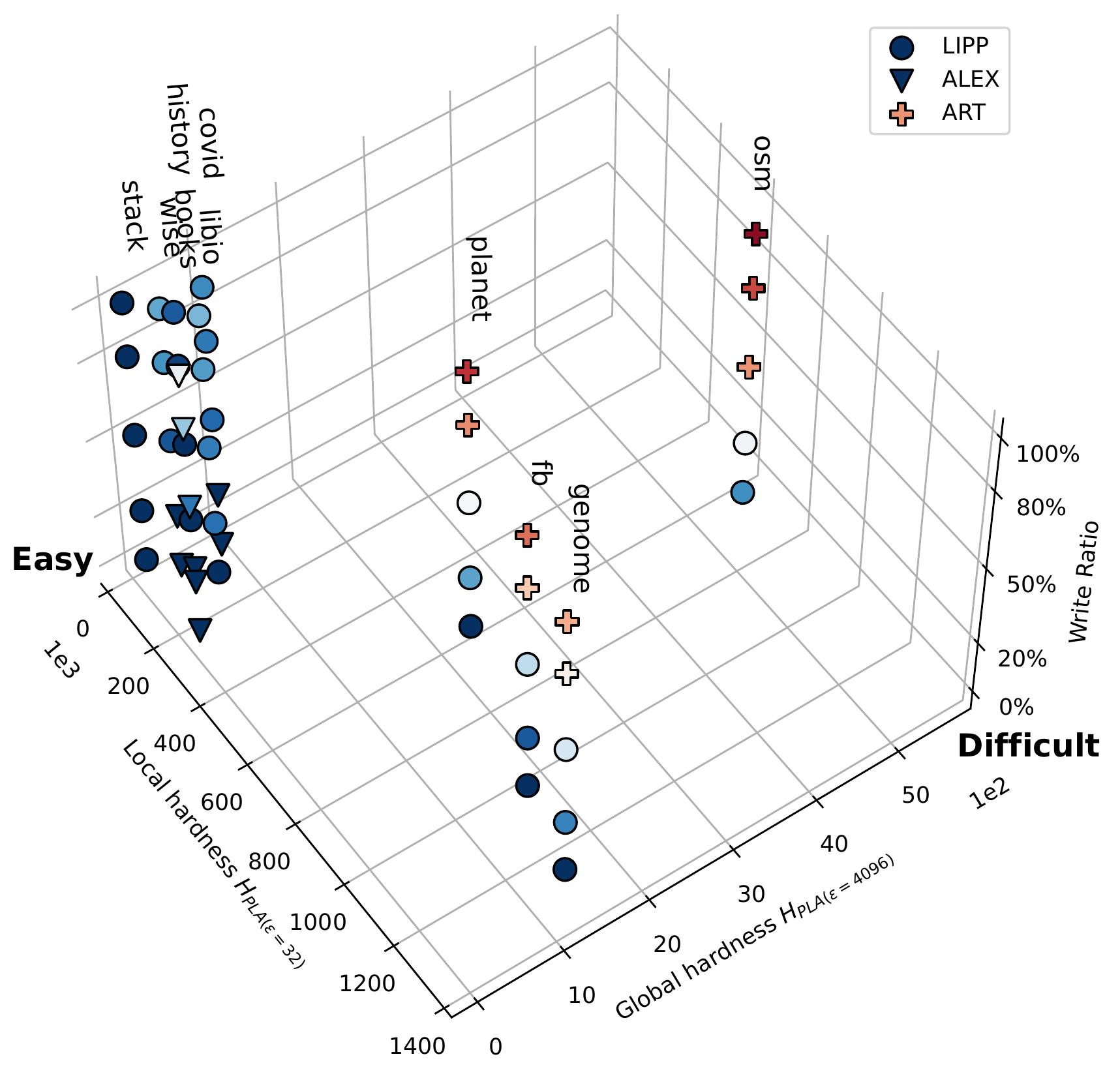}
    \caption{Throughput heatmap (single-threaded) that shows the throughput ratio between the best performing learned index and the best performing traditional index. %
    A positive ratio (in 
    \textcolor{red}{red}) means a \textcolor{red}{traditional index} is the winner under that particular workload and dataset;
    a negative ratio (in \textcolor{blue}{blue}) indicates that a 
    \textcolor{blue}{learned index} is the winner.
    }
    \label{fig:rq1-single}
\end{figure}

\section{Performance}\label{sec:perf}
In this section, we study on the performance of updatable learned indexes in single-core and multi-core settings.

\subsection{Single-Threaded Experiments}\label{sec:singlethread}
We start with single-threaded settings to compare state-of-the-art updatable learned 
indexes with the traditional ones.
Figure \ref{fig:rq1-single} shows a heatmap of \emph{throughput ratios} between learned indexes and traditional indexes. 
The color of each point in the heatmap
indicates the throughput ratio between the \emph{best} learned index 
and the \emph{best} traditional index 
under that specific data-workload. 
A blue point indicates a positive ratio, i.e., 
there exists a learned index that outperforms all the state-of-the-art traditional indexes.
A red point (negative ratio) indicates otherwise, i.e., 
there exists a traditional index that outperforms all the state-of-the-art learned indexes.
The darkness of a point represents the ``winning'' ratio with a darker color indicating the winner outperforms the other indexes by a higher margin. 
Figure \ref{fig:rq1-single} shows that 80\% of the data points are in blue color, leading to our first message:

\begin{msg}
In a single-core environment, 
updatable learned indexes outperform traditional indexes   
over 80\% of our data-workload combinations.
\end{msg}

In fact, we have another four datasets but they are also easy.
To avoid cluttering the easy region,
we include only 10 out of the 14 real datasets.

\begin{msg}
Most real datasets are easy.  
\end{msg}

In Figure~\ref{fig:rq1-single}, %
\alex, \lipp and \art
are the overall winners. 
Strictly speaking, 
\pgm could 
be the winner on 100\% write workloads as it exhibits the highest insert throughput when there is no lookup.
Nonetheless, 
its lookup performance is dominated by ALEX and LIPP; and its good insert performance 
is not attributed to its core design
but its LSM-styled approach to handling inserts.
We therefore do not show \pgm in the heatmap.
The main factors that lead 
\alex and \lipp perform better than the other learned indexes
are (1) the use of ML for subspace lookup in their inner nodes
that eliminate search in inner nodes
and (2) the sparse-node design avoids visiting multiple trees or delta trees as in PGM-Index, XIndex, and FINEdex.
ART as a traditional index outperforms
the other traditional indexes because 
of its cache friendliness \cite{art} and performs especially well on  integer keys (whereas the more recent ones like Wormhole and HOT are specialized for long string keys).
Yet, ALEX and LIPP as learned indexes can outperform ART
except on very hard data
because they are instance-optimized.
Focusing on the write-intensive workloads leads to the third message: 

\begin{msg}\label{msg:hard}
In a single-core environment, 
updatable learned indexes only cannot outperform traditional indexes   
on hard datasets with $\ge$50\% writes.
\end{msg}

While it is known that hard datasets could give a hard time to learned indexes, 
Figure \ref{fig:rq1-single} shows that recent learned indexes like ALEX and LIPP 
have already overcome that on \emph{lookup} operations:

\begin{msg}
In a single-core environment, 
updatable learned indexes outperform traditional indexes from read-only to read-intensive workloads,
regardless of the data hardness.
\end{msg}

Hence, %
with the fact that \emph{lookup is the first step of insert} (an insert of key $k$ would first lookup $k$ to locate the slot to be inserted), what {\bf else} inside a learned index's insert operation
can out-bleed the speed gain from its first step?

\begin{figure}
    \centering
    \includegraphics[width=\linewidth]{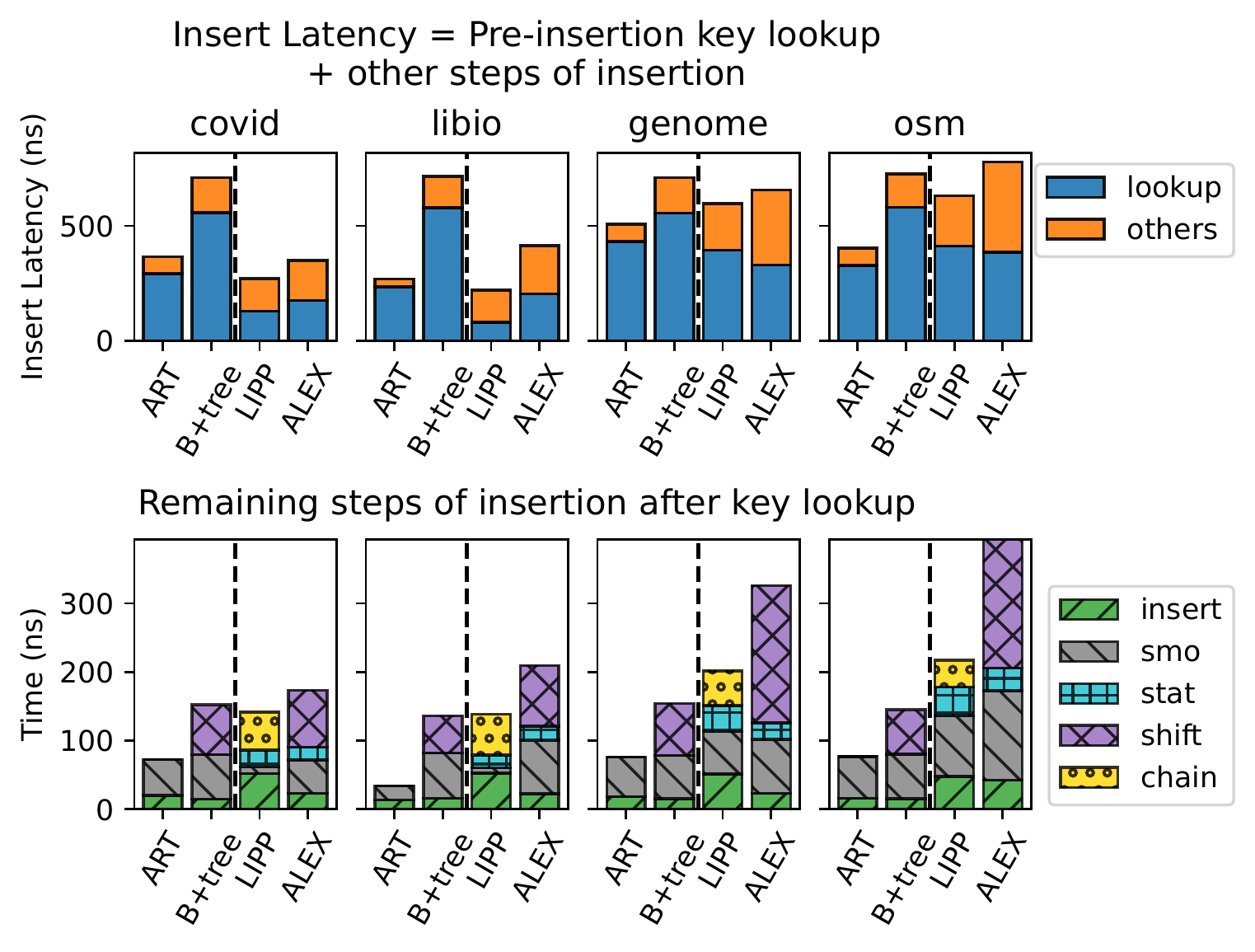}
    \caption{\added{Time breakdown of insert operations.}}
    \label{fig:ins_break}
\end{figure}

To answer this question, 
we break down the average latency of the insertions 
of ALEX and LIPP in the write-only workload.
Figure~\ref{fig:ins_break} shows the result.
We include ART and B+-tree as references.
We show the results for two easy datasets ({\tt covid}) and ({\tt libio}),
the locally hardest dataset ({\tt genome}),
and the globally hardest dataset ({\tt osm}).
Figure~\ref{fig:ins_break}(top) confirms that 
learned indexes generally get a more efficient 
first-step (lookup) in an insertion (except {\tt osm}),
but the remaining steps of an insert 
perform much worse than the best traditional index (\art) and are no better than B+-trees.

Figure~\ref{fig:ins_break}(bottom) details the remaining insert steps, 
where a large part of the latency in learned indexes
is due to collision resolution.
For ALEX, that is the time spent on shifting the elements in data nodes and carrying out SMOs (e.g., node resizing);
for LIPP, that is the time spent on 
creating and chaining new nodes 
and its SMO-like adjustment procedure to bound the tree height.
Although the shifting and SMO costs in ALEX are not specific to learned indexes (also seen in B-tree variants), 
they are more expensive than in B-trees 
and worsen in harder datasets.
In contrast, although \lipp also exhibits higher collision resolution costs 
on harder datasets, it is still smaller than that of ALEX.

\begin{table}
\caption{Statistics of an insert operation in ALEX and LIPP.}
\scalebox{0.85}{
\begin{tabular}{|l|cc|cc|}\hline
Dataset &  \multicolumn{2}{c|}{ALEX} & \multicolumn{2}{c|}{LIPP}  \\ \cline{2-5}
 &  Node traversed & Keys shifted &  Node traversed & Node created \\\hline\hline
{\tt covid} & 1.02 &  8.07 &  1.23 & 0.4 \\
{\tt libio} & 1.04 &  19.92 &  1.09 & 0.4 \\
{\tt genome} & 1.01 & 42.62  & 2.12 & 0.32 \\
{\tt osm} &  1.62 & 35.84 &  2.33 & 0.28 \\\hline
\end{tabular}
}
\label{tbl:ops_count}
\end{table}

Table \ref{tbl:ops_count} further shows the
detailed statistics per insert in ALEX and LIPP.
As the table lists, a harder dataset 
does not particularly increase the number of chaining operations (node creation) in LIPP
but only slightly increase tree traversal time to reach the designated node.
That is because LIPP creates at most one new node on collision, which successfully bounds the write amplification to be one node allocation per collision.
In contrast, ALEX has a number of key shifts and 
the number of shifts increases with the data hardness, because 
a harder dataset challenges ALEX in multiple aspects 
including its cost models, fill factor, and model accuracy. 
ALEX's write amplification (the number of key shifts due to a collision) is large because it is only bound by its huge node size (maximum 16MB).
Figure \ref{fig:ins_break} also reveals a unique component in learned indexes insertion: the update of the various statistics on insertions.  
The cost is non-negligible and is particularly pronounced in \lipp because it updates the statistics in every node on the insertion path.

\begin{msg}\label{lippgoodinsert}
In a single-core environment, 
LIPP's node chaining collision resolution has a lower write amplification than ALEX's key shifting collision resolution.
\end{msg}

With the understanding of where the time goes in insertions,
it seems that LIPP as a learned index is ``ready'' by having 
competitive insert performance and excellent lookup on a majority of real datasets.
However, we observe its design is mainly optimized for single-threaded execution and is often at odds with other important aspects, including multi-core scalability as we discuss next.

\subsection{Multi-Threaded Experiments}

\begin{figure}
    \centering
    \includegraphics[width=0.88\linewidth]{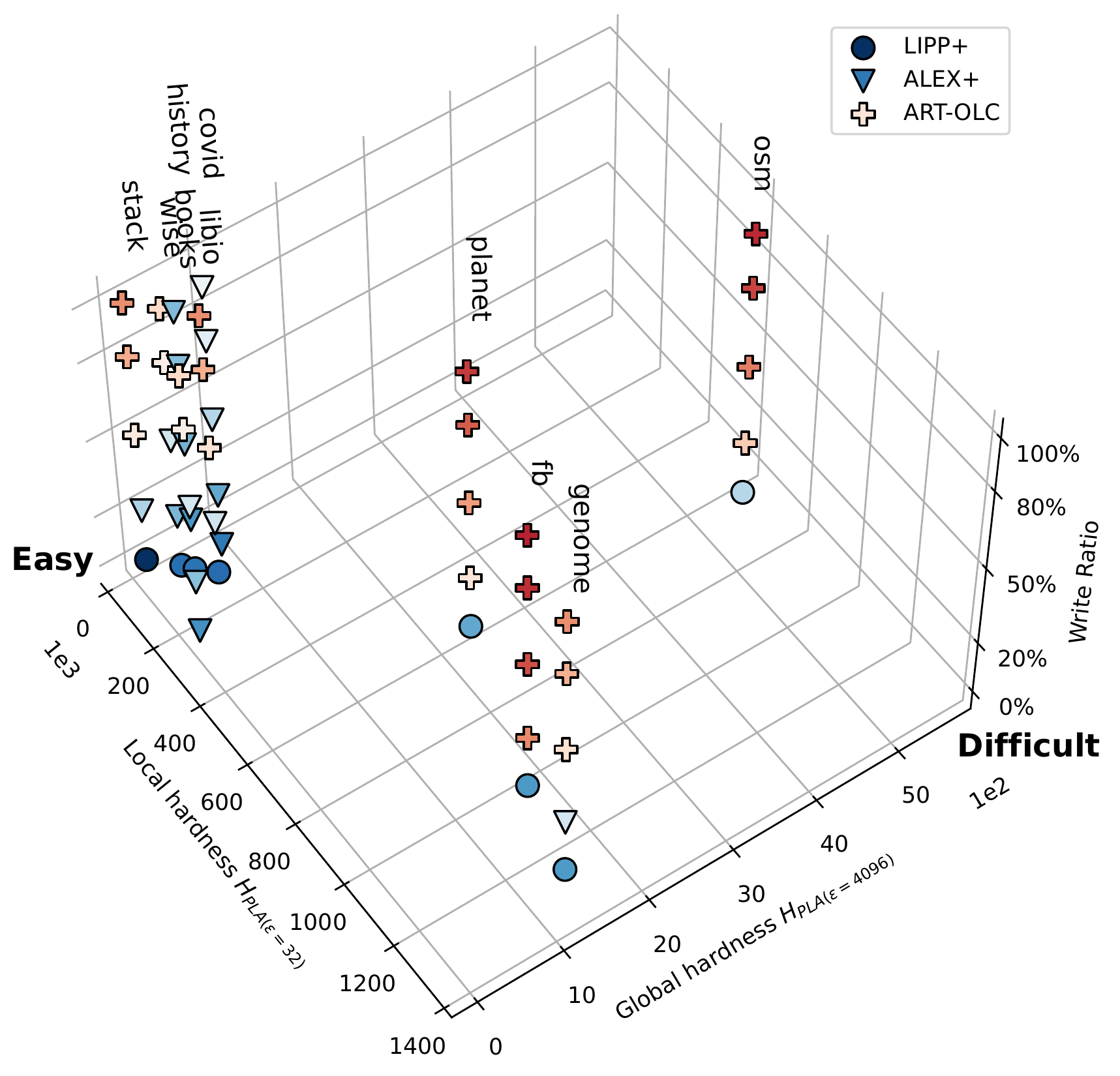}
    \caption{Throughput heatmap under 24 threads (one socket).}
    \label{fig:rq1-multi}
\end{figure}

We now move on to the multi-threading environment and begin with the heatmap that shows the best indexes in our data-workload space under 24 threads without hyperthreading in 
Figure \ref{fig:rq1-multi}. %
LIPP+, ALEX+ and ART-OLC are the only winners.
However, two notable changes are that 
LIPP+ has lost its leading position to ALEX+ except on read-only workloads;
and 
ART-OLC also has taken over some easy datasets on write-intensive workloads.
The latter is because ART has been very competitive --- 
sometimes its performance is close to LIPP and is better than ALEX in the single-threaded setting.
When LIPP+ loses its edge in the multi-threading setting,
ART-OLC takes over as the best performing index, followed by \alexplus.

\begin{figure*}
    \centering
    \includegraphics[width=0.88\linewidth]{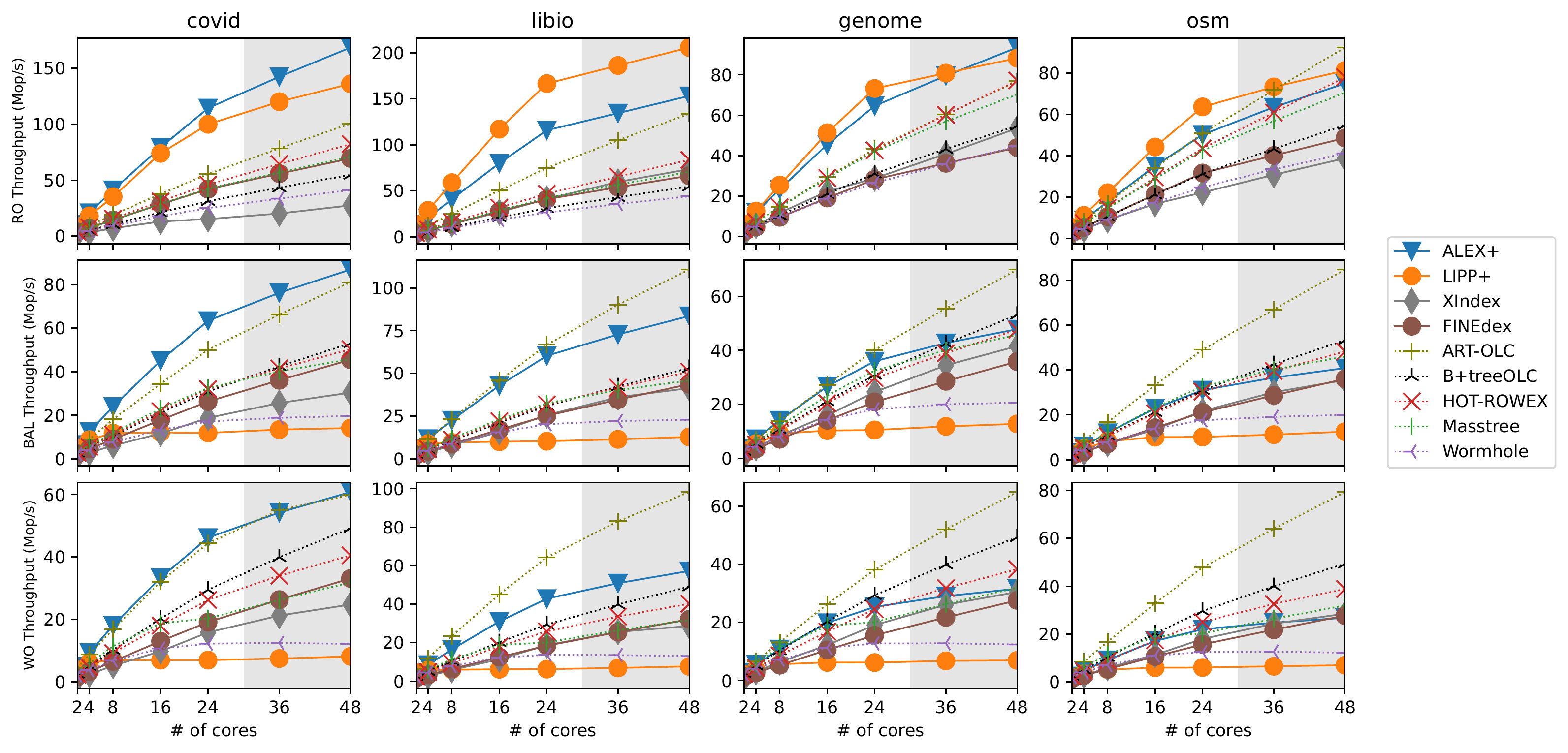}
    \caption{Throughput of the read-only (top), balanced (middle) and write-only (bottom) workloads using one socket. The grey area indicates hyper-threading is enabled.}
    \label{fig:scalability}
\end{figure*}

Figure \ref{fig:scalability} (white area) shows the scalability from 2 to 24 cores of the read-only, balanced, and write-only workloads. 
All the learned indexes scale well on read-only workloads.
However, once the workloads include writes,
LIPP+ can no longer sustain its scalability despite its use of optimistic concurrency because every insert thread has to update the per-node statistic on its insertion path.
That induces high contention and cacheline ping-pong, especially at the root node. 
This is a  drawback of using a unified node layout where statistics have to be maintain in every node rather than in only the leaf nodes like ALEX+.
In contrast, ALEX+ scales well by taking the advantage of lock-free lookup and optimistic locking in leaf nodes, until  %
its write amplification becomes severe under hard data.

\begin{msg}
After parallelization, some single-threaded updatable learned indexes (e.g., \alex) 
can scale and perform better than native concurrent learned indexes (e.g., \xindex), but some (e.g., \lipp) cannot and perform worse.
\end{msg}

\subsection{Impact of Hyper-threading/NUMA}

The grey area of Figure \ref{fig:scalability} shows the scalability with 36 and 48 threads with hyper-threading (48 hardware threads per socket).
When writes are involved, LIPP+ cannot scale and Wormhole's single lock for the inner layer severely limits concurrency. 
Except them, all indexes benefit from hyper-threading but show different degree of performance saturation.
For example, \alexplus exhibit slow down from memory bandwidth exhaustion due to its high write amplification and long last-mile search, especially on harder data. In fact, our profiling results indicate that ALEX+ has already saturated the memory bandwidth with 24 threads in one socket.

\begin{figure*}
    \centering
    \includegraphics[width=0.88\linewidth]{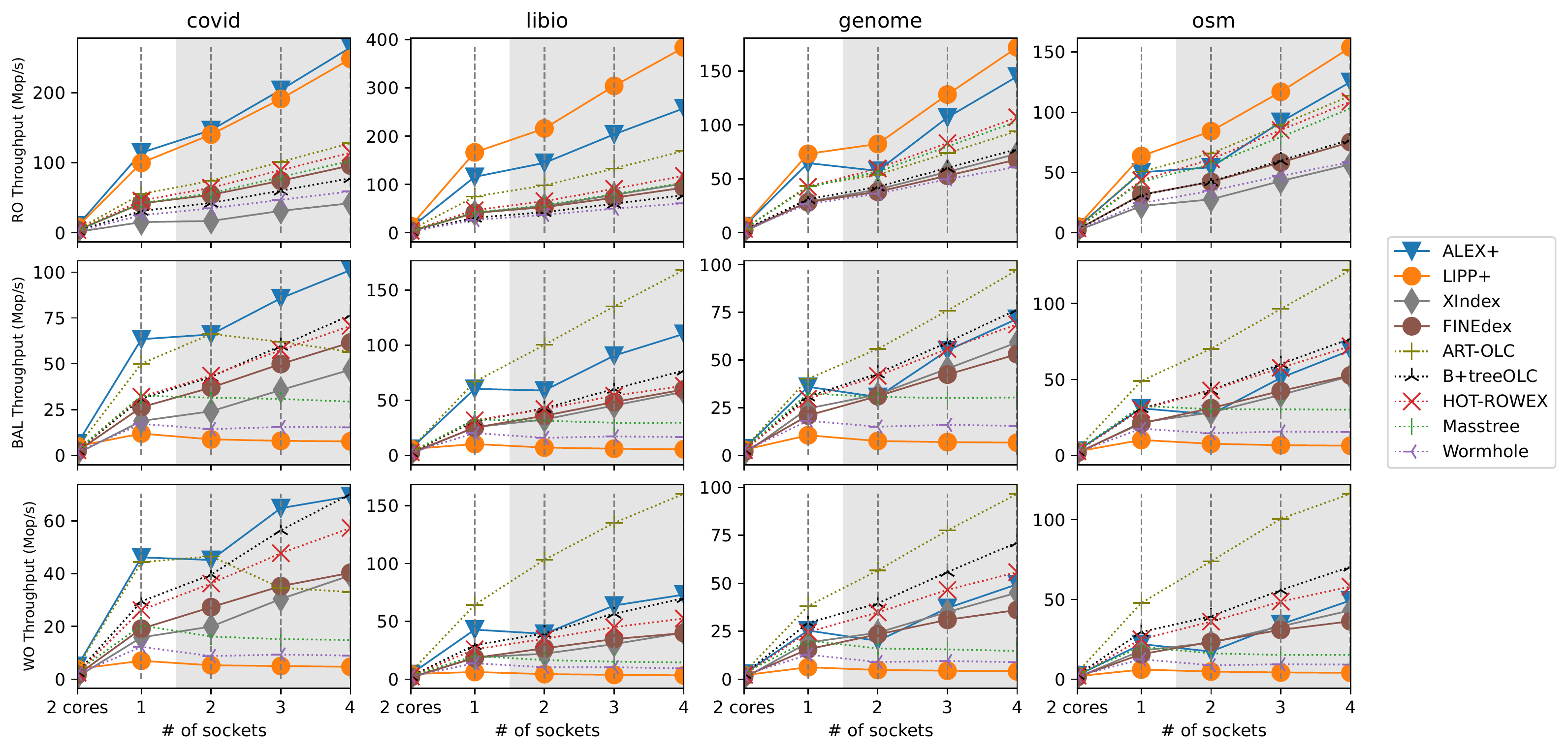}
    \caption{Throughput under varying socket counts.}
    \label{fig:scalability_numa}
\end{figure*}

Moving on to NUMA, Figure \ref{fig:scalability_numa} shows the scalability of each index when scaling from 1 to 4 sockets. 
We use the {\tt Interleave} NUMA memory allocation policy that allocates memory pages from different sockets in a round-robin fashion. This is also the setting which yields the best performance for all indexes.

All indexes show diminishing return
once we use more than one socket due to 
cross-socket bandwidth is lower than the intra-socket (24 cores) bandwidth.
ART-OLC has difficulty scaling on easier datasets ({\tt libio} and {\tt covid}). That is because those datasets have dense keys that result in dense nodes and hence, greatly increase contention on high thread counts. 
Masstree crumbles when there are writes 
because its write amplifications and concurrency control 
together have exhausted
the cross-socket bandwidth~\cite{hydralist}. 
ALEX+, performs worse either with insertion or harder data on two sockets because distributing the memory accesses over two sockets 
would experience the tighter cross-socket bandwidth bottleneck despite the increased bandwidth in aggregation.
ALEX+ however can scale again with more sockets as by then there are more bandwidth channels.
Overall, this experiment shows that:

\begin{msg}\label{msgnuma}
Hyper-threading and NUMA have influence on the
scalability of learned indexes to various degree,
and their core designs play a role in their scalability under hyper-threading/NUMA.
\end{msg}

\subsection{Deletion Performance}
\begin{figure}
    \centering
    \includegraphics[width=0.88\linewidth]{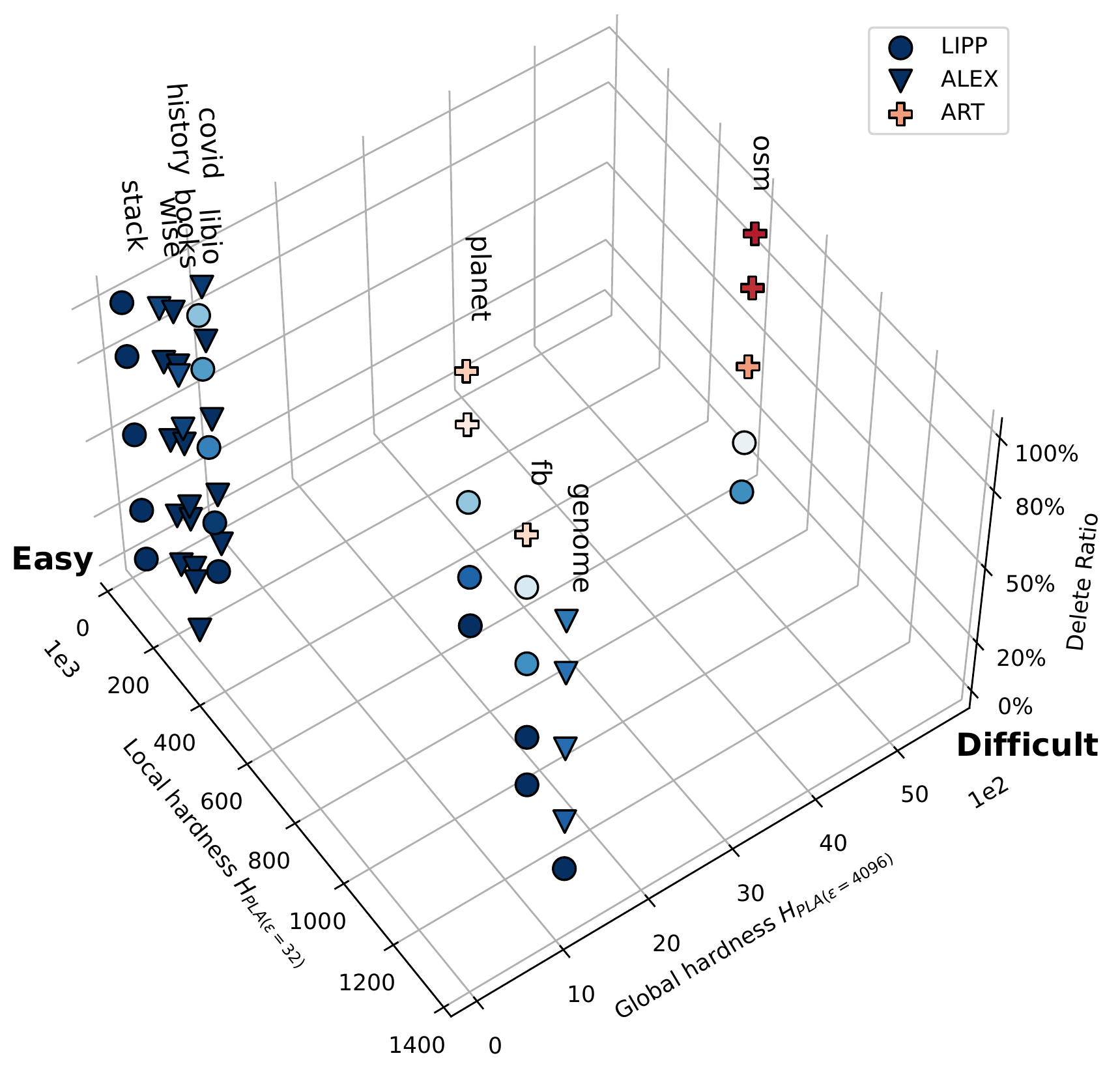}
    \caption{Throughput heatmap (single-threaded) under deletion workloads.}
    \label{fig:delete}
\end{figure}

Before we move on to the next section,
we complement this section using deletion workloads.
LIPP, Masstree, Wormhole, B+TreeOLC, and HOT-ROWEX do not cover deletions.
We exclude them from our study except for LIPP (and LIPP+),
which we implemented deletion for it.
We also extended ALEX's deletion for \alexplus.

In this experiment, 
for each dataset, we bulk load all the 200M keys,
and randomly issue lookup/delete requests until 100M keys are deleted.
Figure~\ref{fig:delete} shows the single-threaded throughput heatmap under five different deletion workloads, from read-only to delete-only (100\% delete).
The results of multi-threading are similar and are omitted for space.
From the results, we see that ALEX, LIPP, and ART (as well as their concurrent versions) are still the most competitive ones.
Note that when compared with the insertion workloads (Figure \ref{fig:rq1-single}),
the learned indexes take over more territory from ART even on hard data.
Although the deletion path of a learned index is analogous to its insertion path, where
deleting a key may also incur write amplification (filling up the gaps) and trigger SMOs (node resizing), 
one crucial difference is that deleting a key from a data node 
would not ``pollute'' a node's ML model.
Consequently, deletion and lookup could continue to enjoy high quality model-based search 
without model pollution or retraining overhead.
This makes learned indexes perform even better and outperforms traditional indexes on more datasets and workloads.

\begin{msg}\label{msgdelete}
Deletions in learned indexes are lightweight
because there is no model pollution.
\end{msg}

\section{Memory Space Efficiency}\label{sec:space}

Most previous work on learned indexes excluded the size of the leaf layer when evaluating memory space efficiency.
We aim to study the end-to-end space consumption (i.e., the size of the whole index including both non-leaf and leaf layers) of the indexes.
We report the size of the indexes 
after running the write-only
workload where the first 100M keys 
are bulk loaded and the rest of them are individually inserted into the index.

\begin{figure}
    \includegraphics[width=\linewidth]{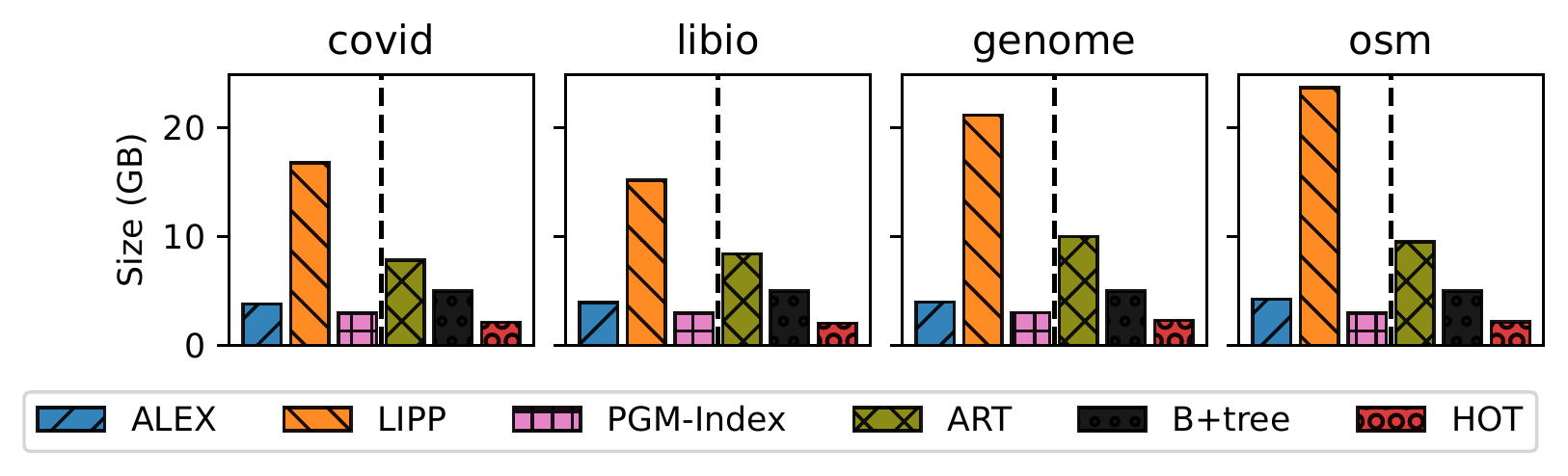}
    \caption{Memory space efficiency of the best indexes.  }
    \label{fig:memory}
\end{figure}

Figure \ref{fig:memory} shows the end-to-end space consumption of the indexes.
For clarity, we present only indexes that are the best either in terms of throughput or space.
From the figure, we observe that:

\begin{msg}
When considering the index size end-to-end, 
updatable learned indexes only have at most 3.2$\times$ space saving over state-of-the-art traditional indexes, and all of them use even more space than HOT.
\end{msg}

The factor 3.2$\times$ 
comes from measuring the difference between the sizes of the most space-efficient learned index (\pgm) and the least space-efficient traditional index (\art). 
Although learned indexes do have small non-leaf layers, memory-optimized traditional index like HOT
can also be very space-efficient.

Although \pgm and \hot are space-efficient, they did not show outstanding performance.
Concerning the three winners in the performance heatmaps,
ALEX is the most space-efficient because its inner layers store no keys.  
Yet it is slightly larger than \pgm because it leaves gaps for insertions.
ART as a trie, has low space utilization in its nodes 
(many null child pointers) when the key space is not dense enough~\cite{hot}.
Despite its excellent single-threaded performance, 
LIPP's memory consumption is 4--5$\times$ larger than ALEX's, and is the highest in this experiment.
Having this observation, one might wonder (1) \textit{what would the  performance of LIPP and ALEX be if 
they were given the same memory space?} and 
(2) \textit{is LIPP's single-thread advantage a fundamental achievement or more a space-performance tradeoff?}

\begin{figure}
    \centering
    \includegraphics[width=0.9\linewidth]{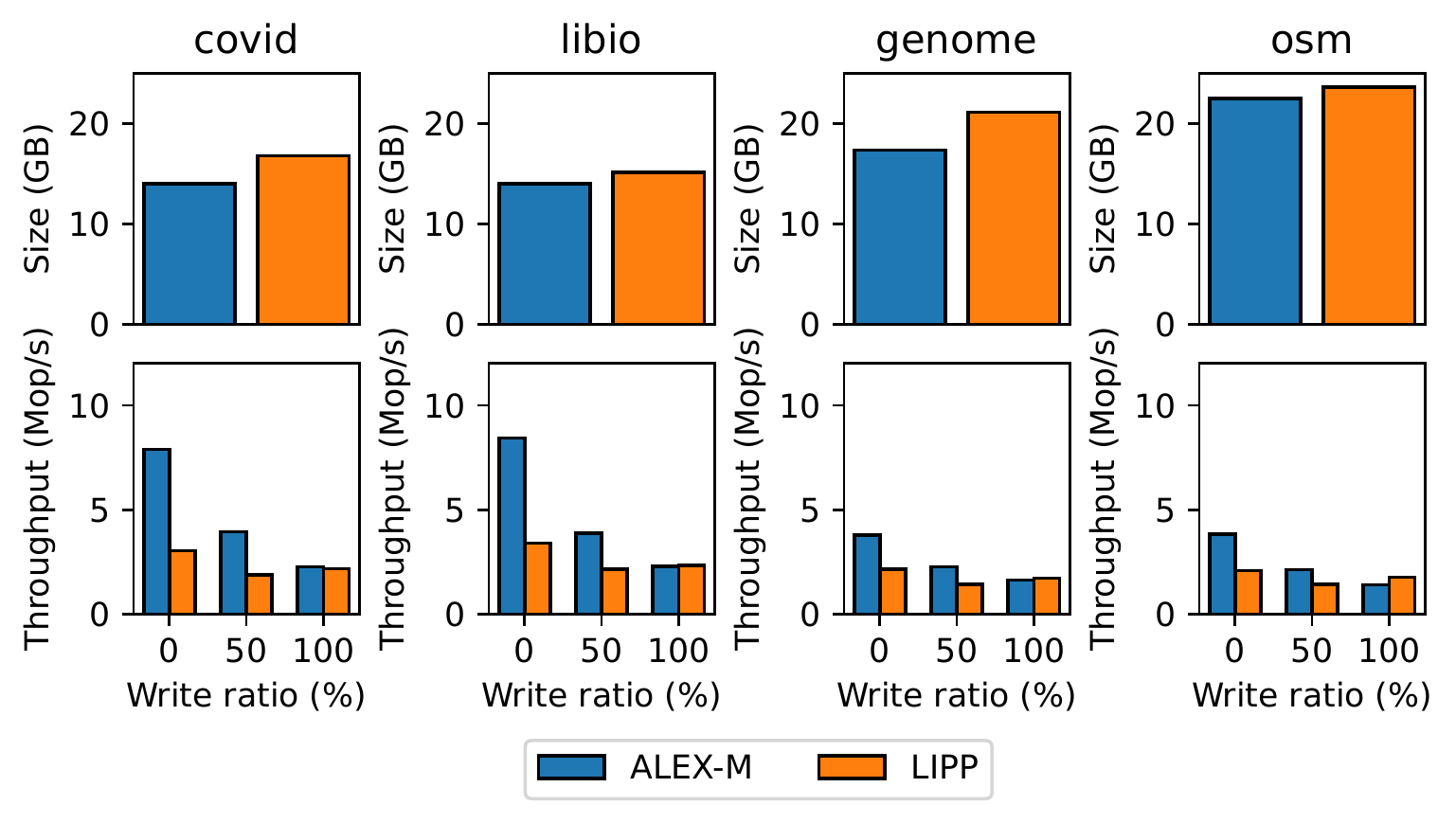}
    \caption{ALEX-M vs. LIPP (when ALEX is tuned to use roughly the same amount of memory as LIPP).}
    \label{fig:bigger_alex}
\end{figure}

\begin{figure*}[t]
    \centering
     \begin{subfigure}{0.45\linewidth}
         \centering
         \includegraphics[width=\linewidth]{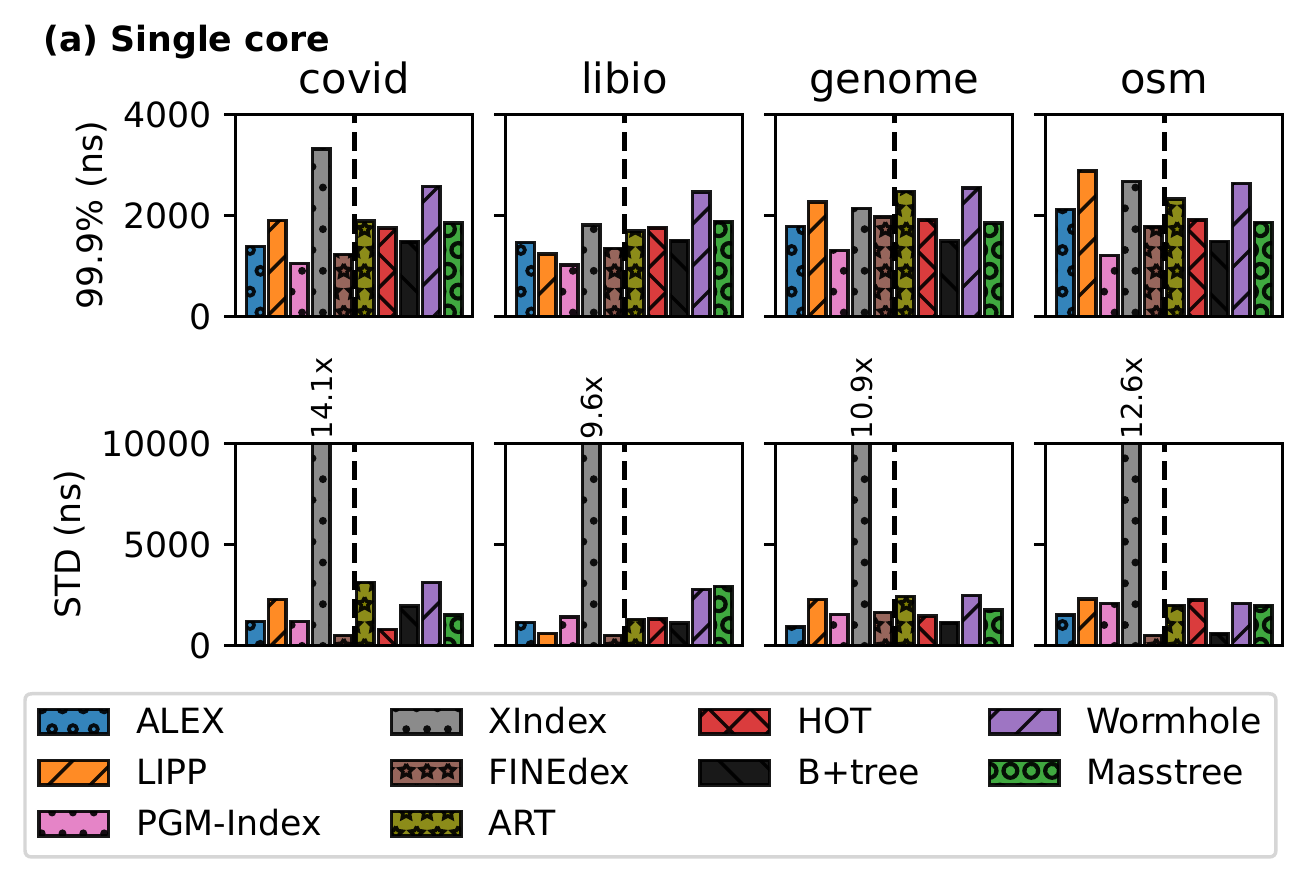}
     \end{subfigure}
     \begin{subfigure}{0.45\linewidth}
         \centering
         \includegraphics[width=\linewidth]{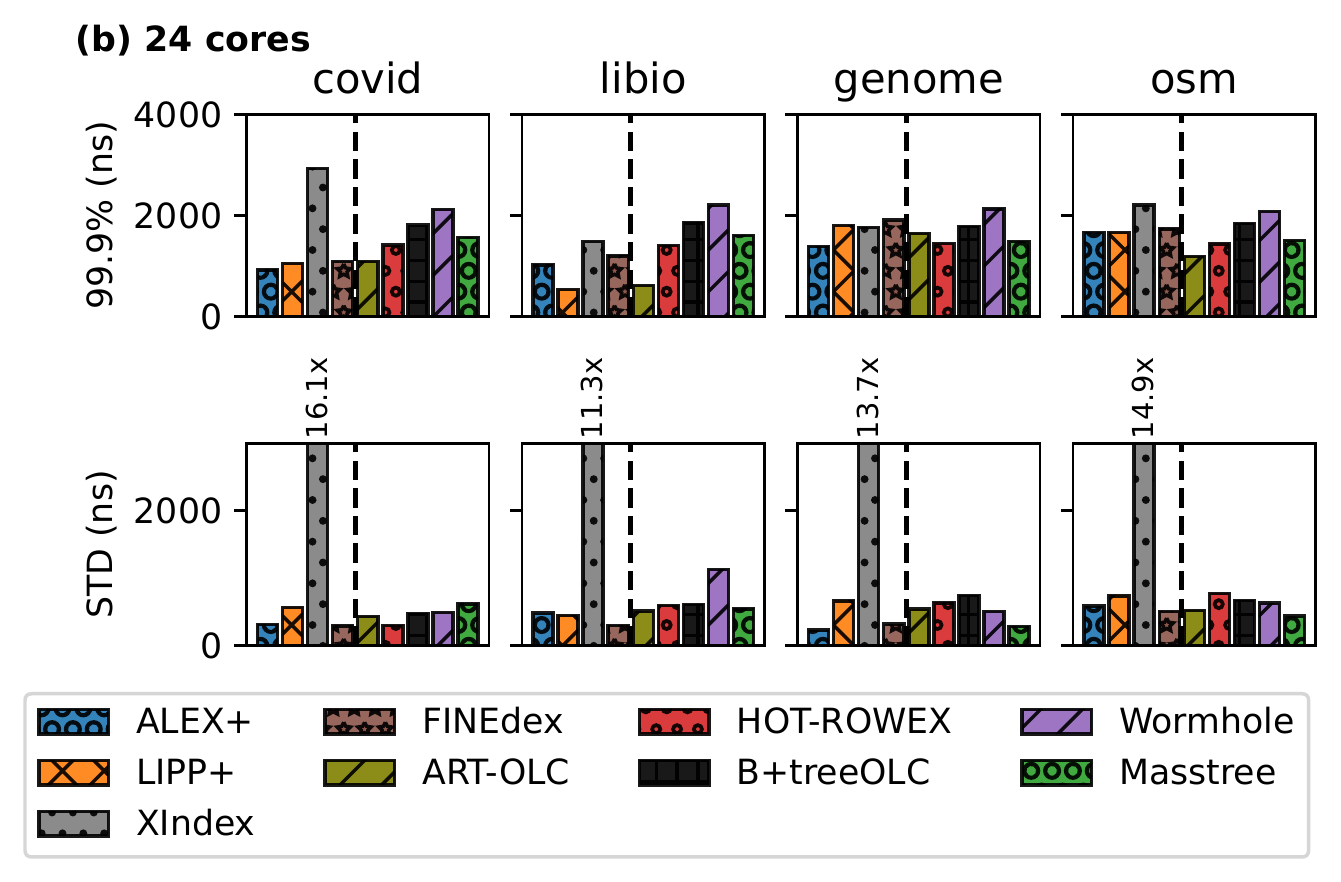}
     \end{subfigure}
    \caption{Tail latency of lookup operations.}
    \label{fig:latency-s}
\end{figure*}

\begin{figure*}
    \centering
     \begin{subfigure}{0.45\linewidth}
         \centering
         \includegraphics[width=\linewidth]{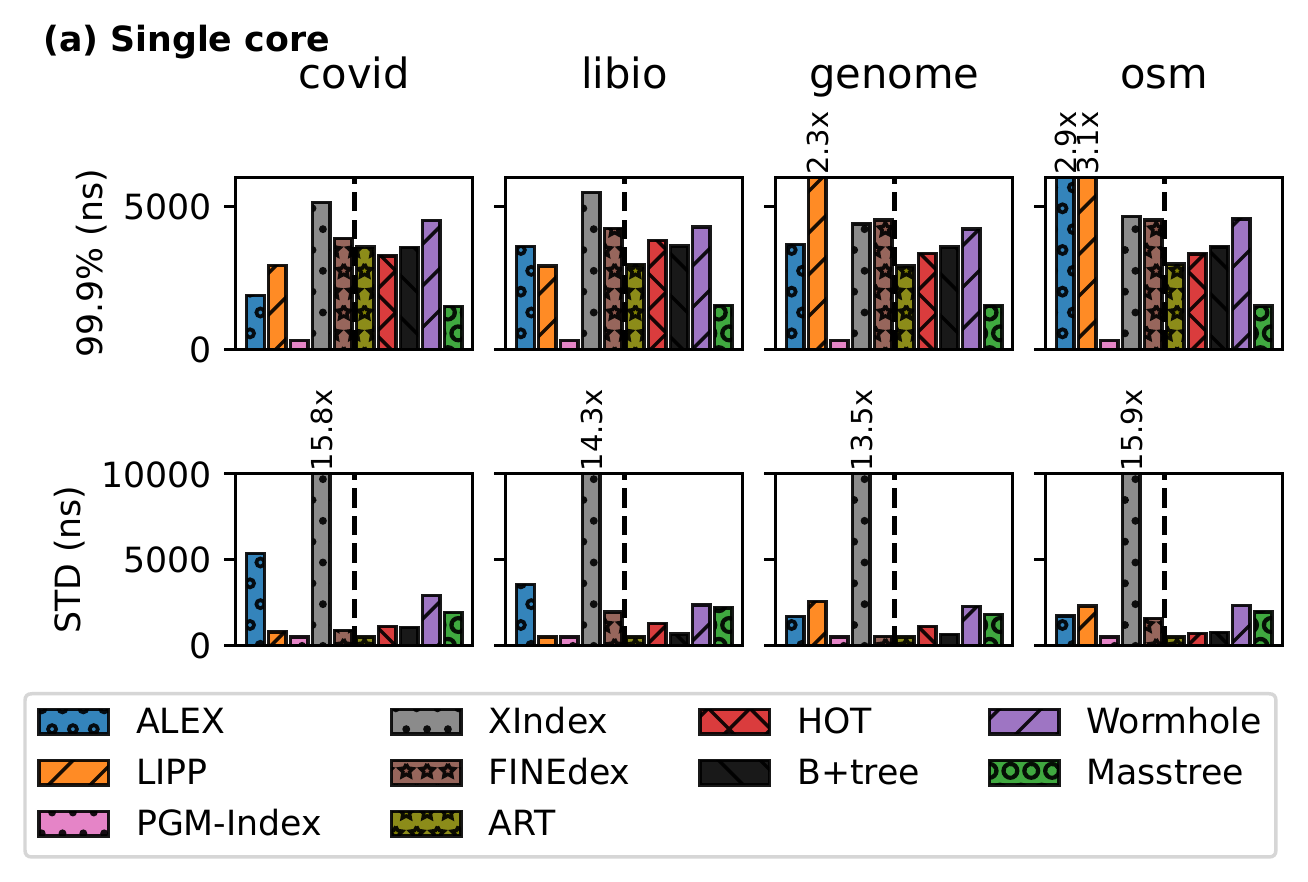}
     \end{subfigure}
     \begin{subfigure}{0.45\linewidth}
         \centering
         \includegraphics[width=\linewidth]{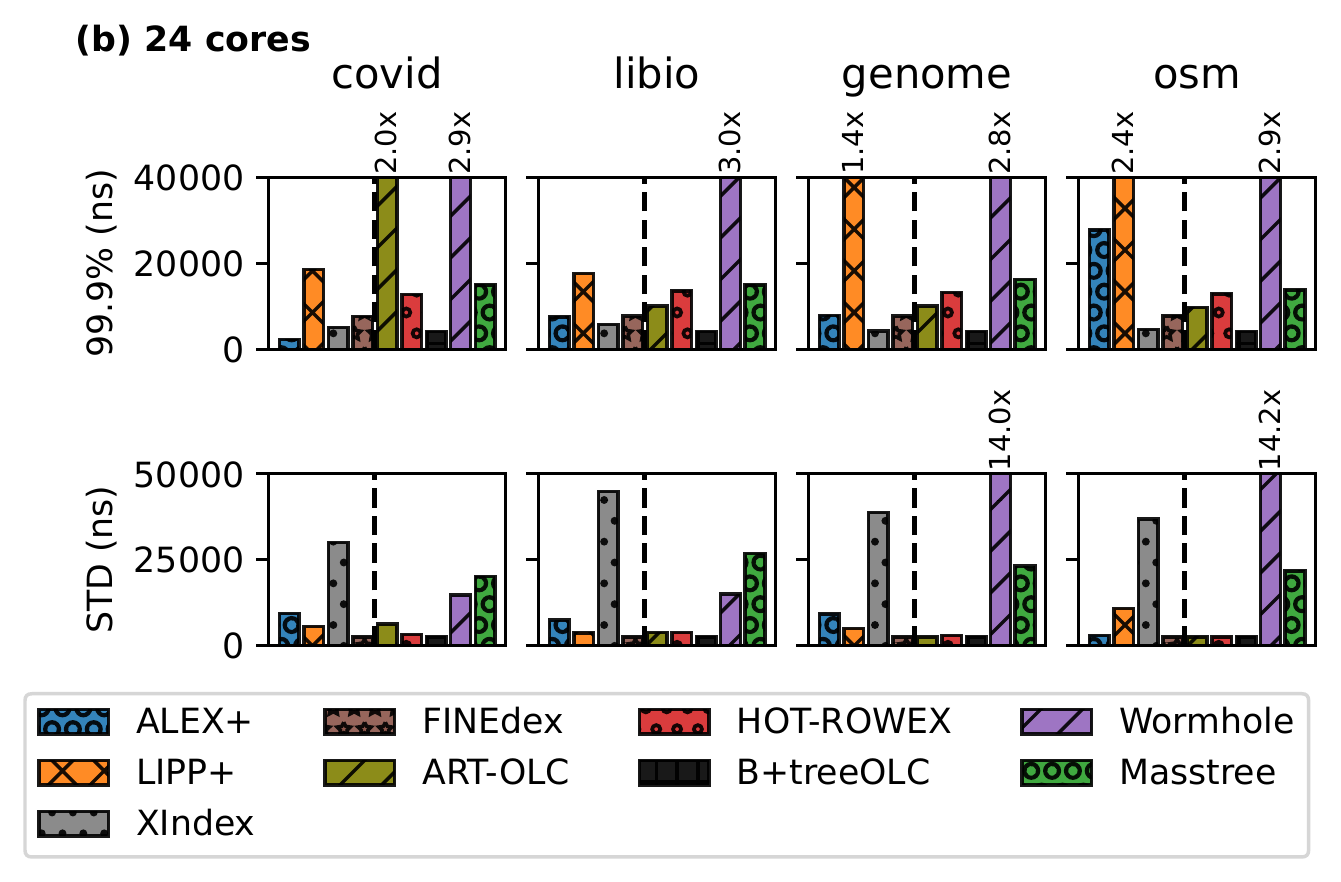}
     \end{subfigure}
    \caption{Tail latency of insert operations.}
    \label{fig:latency-i}
\end{figure*}

To answer these questions, we carried out another experiment 
which tuned the fill factor of ALEX data nodes such that the resulting index (ALEX-M) uses roughly the same amount of memory space as LIPP.\footnote{We cannot do the other way round because of LIPP's implementation sets its fill factor as an integer;
hence we could not set any fill factor smaller than 2, and a fill factor of 1 means no gap is allowed.}
The resulting fill factor of ALEX-M is 0.2--0.25, 
whereas ALEX has an original fill factor of 0.7. 
Figure \ref{fig:bigger_alex} shows the throughput of ALEX-M and LIPP under this new setting.  %
ALEX's lookup performance has improved significantly and dominated LIPP on both easy and hard datasets. 
The reason is that with lower density data nodes, an insert in ALEX can often find a gap, maintaining high model accuracy and incurring fewer key shifting. 
These results indicate that \lipp's collision-driven design is trading space for speed.

\section{Robustness}\label{sec:robust}
A robust index shall have low tail latency
despite any concurrency degree or 
heavy-lifting part (e.g., SMO).
A reliable index shall also have a robust performance as
the underlying data distribution changes. 
Furthermore, %
it is important for the indexes to perform robustly under a variety of range scan sizes. %

\subsection{Tail Latency}
In this experiment, 
we report the tail latencies (variance and 99.9 percentile) of the indexes 
under both single-threaded and multi-threaded (24 cores) settings. 
The lookup and insert latencies are sampled from 1\% of the operations from the read-only and write-only workloads, respectively.
As Figure \ref{fig:latency-s}a shows, the updatable learned indexes (except \xindex) exhibit comparable tail latency
with the traditional indexes
under single-threaded lookup operations. 
XIndex's latency variance is especially high due to 
the expensive context switching between its foreground and background threads.
Specifically, unlike other indexes,
XIndex requires an extra background thread to merge deltas. 
For fair comparison, 
we pinned its operational and background threads to the same core so that all the indexes are evaluated using the same CPU budget. 
This experiment reveals that using 
background threads to handle dynamic workloads would hurt latency variance.
We have confirmed this reason
by pinning the background thread to an extra physical core, after which its tail latency is back to normal.
The results using multiple threads (Figure \ref{fig:latency-s}b)
are similar.
It is also worth-noting that LIPP+'s tail latencies remain low even under multi-threading, although it did not scale well on throughput. 
The reason is that LIPP+ uses atomic instructions to update statistics, affecting average latency rather than tail latency.

\begin{figure}
    \includegraphics[width=\linewidth]{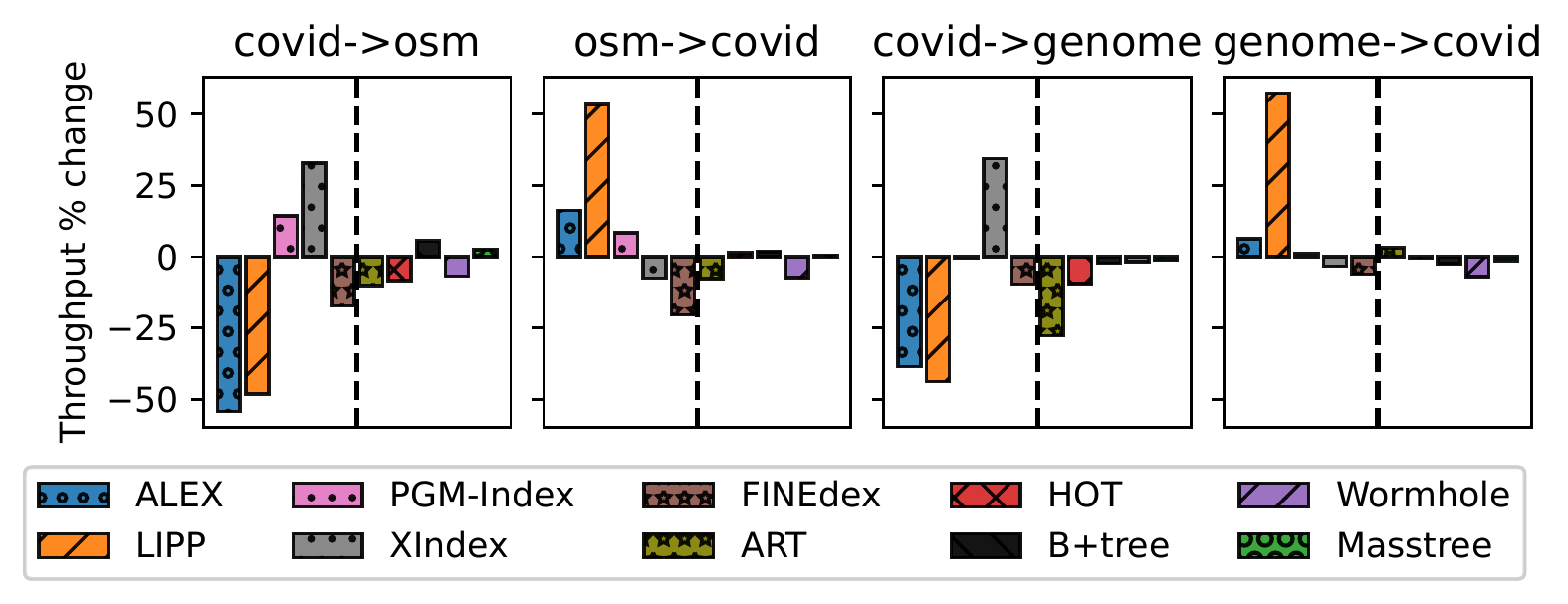}
    \caption{Throughput changes as data distributions change.} 
    \label{fig:datashift}
\end{figure}

Figure~\ref{fig:latency-i} shows the tail latency for inserts. 
With a single thread (Figure \ref{fig:latency-i}a),
updatable learned indexes except XIndex generally have similar tail latencies
as the traditional ones.
ALEX and LIPP are sensitive to data hardness.
They have high 99.9\% tail insert latency on the hard {\tt osm} and {\tt genome} datasets due to the increased number of SMOs. %
Similar to its lookup operations, XIndex's insert operations exhibit very high variance in tail latencies regardless of the data hardness because of the context switching overhead between its 24 foreground threads 
and 3 background threads using 24 physical cores.\footnote{The foreground and background thread ratio follows the recommendation from XIndex's original paper~\cite{xindex}.}
Overall, the multi-threaded results are similar to the single-threaded ones,
except that Wormhole has a higher tail latency for inserts 
due to its use of a single exclusive lock for the whole inner layer.

\begin{msg}
Except XIndex, updatable learned indexes exhibit low 
tail latency on both single- and multi-threaded settings.
Yet, some traditional indexes (ART, HOT and B-tree) exhibit impeccable robustness in this aspect.
\end{msg}

\subsection{Shifting Data Distributions} 
The goal of our next experiment is to study how learned indexes 
behave and adapt to changes of data distribution after the index has been deployed.
We follow earlier work~\cite{alex} to 
(1) bulk load an index using 100 million keys 
in one dataset $X$,
and (2) start a read-write balanced workload using 100 million keys of another dataset $Y$ for insertions and lookups for the keys in $X$. 
The keys of both datasets are scaled to the same domain.
With the ability to quantify data hardness, 
we shift from easy data ({\tt covid}) to two different kinds of hard data ({\tt genome} and {\tt osm}) and vice versa.

Figure~\ref{fig:datashift} shows the change of the throughput on the balanced workload with respect to the original workload with no change of the dataset.
The result shows that learned indexes are sensitive to data distribution changes while traditional indexes are not.
The changes, however, can be both positive and negative. 
For example, ALEX's throughput can drop by up to 52\% 
when it is bulk-loaded with easy data {\tt covid}, followed by inserts of hard data ({\tt osm}).
However, its performance can improve by up to 15\% 
when the hard data {\tt osm} is first bulk-loaded, followed by the easier {\tt covid}. 
The result aligns with the observations made by prior work~\cite{alex},
where the index starts with easy data can incur significant overhead to
adapt to the new, harder distribution.
Yet, an index started with harder data requires less/no overhead to adapt to easy data.
LIPP behaves similarly to ALEX.
\pgm and \xindex are more resilient to distribution changes.
In \pgm, the different distributions are likely to be stored in different trees in its LSM structure.
\xindex's throughput is less sensitive to the increased SMOs as they are handled by background threads,
which impact tail latency.

\begin{msg}
Learned indexes are sensitive to data distribution changes, while traditional indexes are not.
Corroborating with prior work, it is harder (easier) for a learned index that is pre-filled with easy (hard) data to adapt to a harder (easier) dataset.  
\end{msg}

\subsection{Range Queries} 
\begin{figure}
    \centering
    \includegraphics[width=1\linewidth]{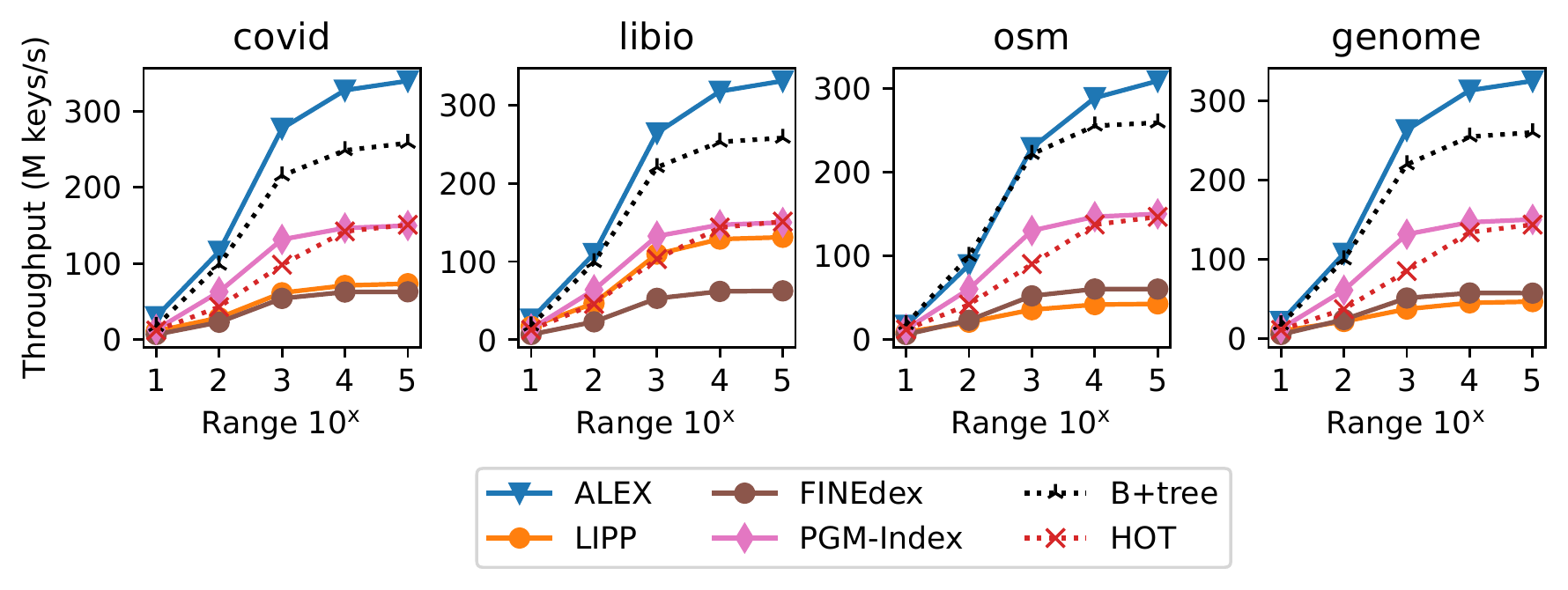}
    \caption{Range query throughput under varying scan sizes.}
    \label{fig:range}
\end{figure}

This experiment evaluates range query performance. 
We bulk load each index using the whole dataset of 200M keys 
and start a read-only scan workload.
Each query picks a random start key $K$
and fetches a fixed number of keys starting from $K$.
Each workload issues 10 million range queries in total
and we measure the throughput in number of keys accessed per second.
Figure~\ref{fig:range} shows the results (for indexes that implemented range scan only) under a varying range query size
from 10 to 10,000 under a single thread. 
As shown by the figure,
all indexes exhibit higher throughput as the query size increases 
because a larger scan size involves less tree traversal
and more efficient in-node scan.
However, this experiment also reveals yet another drawback of LIPP's unified node design: with a node layout that interleaves child pointers and data in the node array, a range scan on the array would inevitably encounter a lot of branches.  
Specifically, to continue to the next entry in the data array,
LIPP needs a branching instruction to decide whether the entry is a child entry 
(hence recursively visiting the subtree) or a data entry.
This largely cancels out the potential improvement brought by using fewer traversals under larger scan sizes. %

\begin{msg}
Learned indexes are good on range queries.  
Yet, the unified node design is not range-scan friendly.
\end{msg}

\section{Complementing Real Data with Synthetic Data}\label{sec:syn}

\begin{figure}
    \centering
\includegraphics[width=0.88\linewidth]{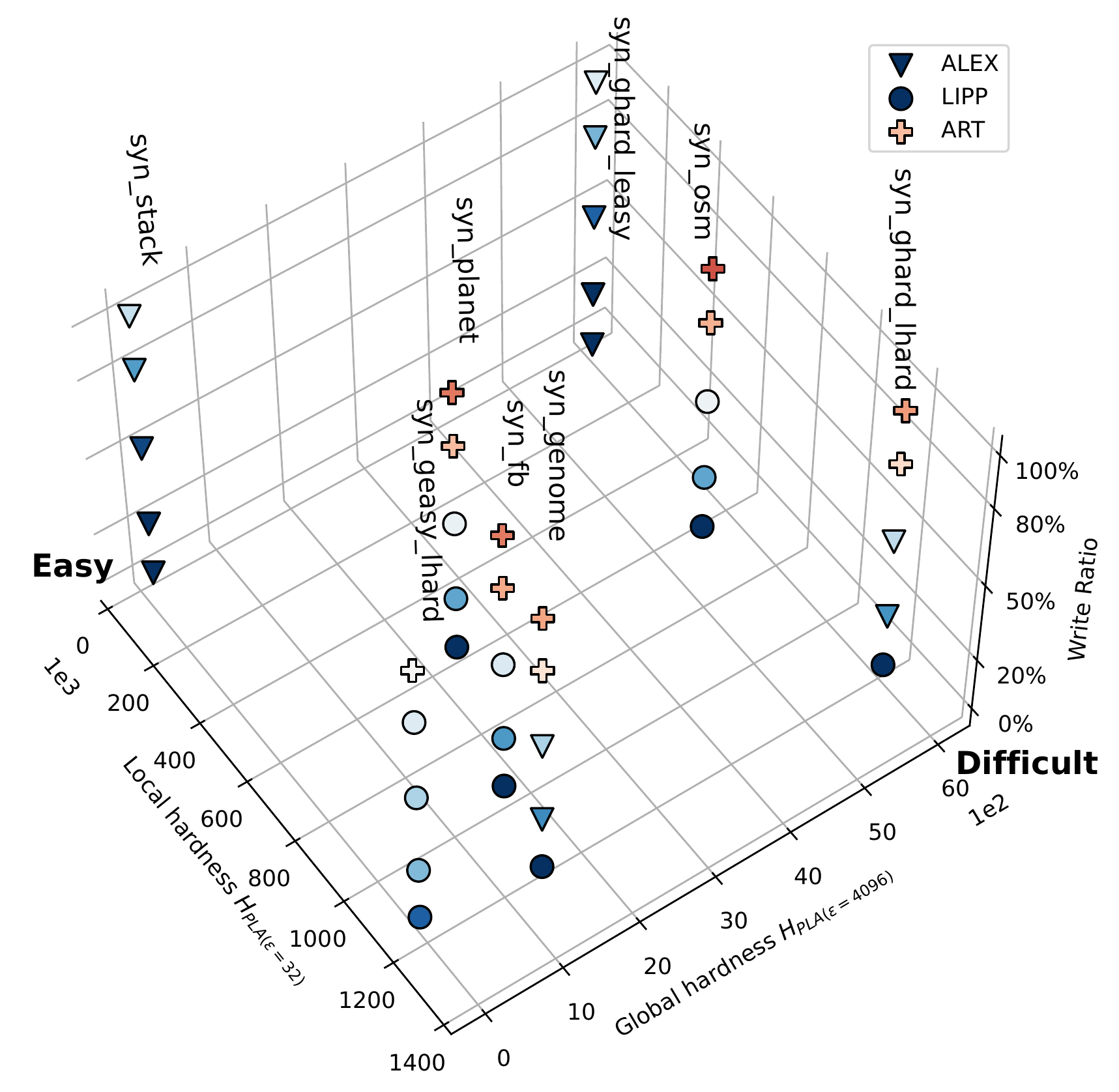}
    \caption{Throughput heatmap (single-thread) from synthetic datasets.}
    \label{fig:syn}
\end{figure}

\begin{figure}

\begin{subfigure}{0.32\linewidth}
    \includegraphics[width=\linewidth]{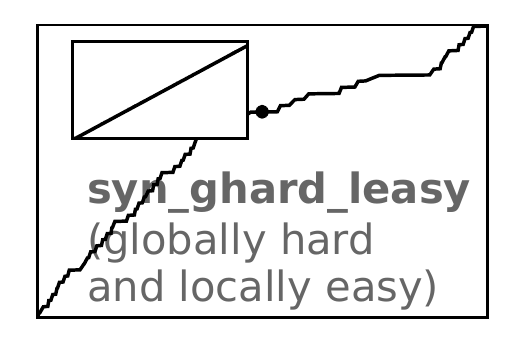}
\end{subfigure}
\begin{subfigure}{0.32\linewidth}
    \includegraphics[width=\linewidth]{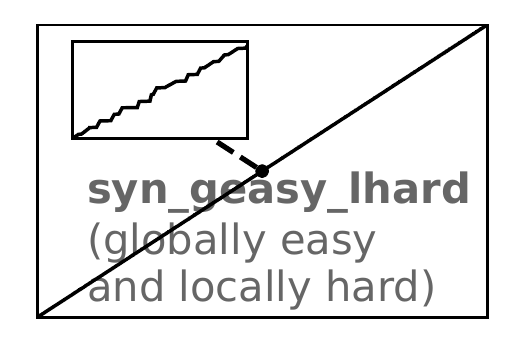}
\end{subfigure}
\begin{subfigure}{0.32\linewidth}
    \includegraphics[width=\linewidth]{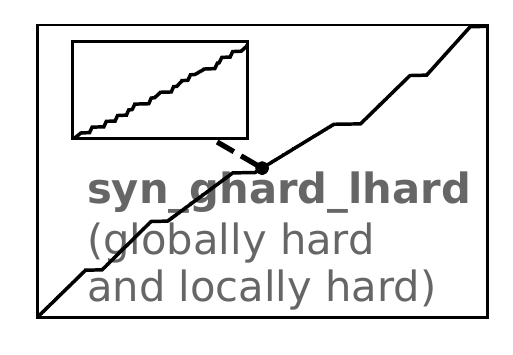}
\end{subfigure}

\caption{Synthetic datasets from heatmap corners.}
    \label{fig:syncdf}
\end{figure}

The global and local hardness defined based on PLA are mildly correlated because they are extracted from the same CDF after all, just in different granularity.
This also explains why we seldom find real datasets 
that are positioned near the ``hard'' corners 
(e.g., globally-hard-locally-easy) 
in our heatmaps. 
Although we find no real dataset that exhibits those extreme hardness,
we have built a synthetic data generator that generates data using local and global hardness as inputs.  
Figure \ref{fig:syn} is the corresponding single-thread throughput heatmap based on the generated data.
We selected local/global hardness values $H$
from the ``hard'' corners of the heatmap.
Figure \ref{fig:syncdf} illustrates their CDFs.

Our synthetic data generator samples data from
a set of random linear models.
Specifically, we first randomly generate a positive slope $m$ and an intercept $b$ for a segment's linear model.
Given that a linear model maps keys to positions, we have $y=mk+b$, where $k$ is a key and 
$y$ is its rank.
For a given rank and $\epsilon$,
a key is then uniformly sampled from $[max(\frac{y-\epsilon-b}{m}, prev + 1), \frac{y+\epsilon-b}{m}]$, where $prev$ 
is the key from rank $y - 1$. 
Keys are incrementally generated from rank 1 to rank 200M.
A segment size is controllable but 
currently we simply give each segment an equal number of keys.
Following the algorithm we described in \cite{gre} for computing data hardness,
we create a convex hull for all keys of a generated segment. 
Then, we generate the first key $k_i$ of the next segment
by incrementing the value of $k_i$ 
until $(k_i, r_i)$ goes beyond 
the bounding box of the convex hull of the previous segment.
The process is iterative and recursive.  
We first generate a global segment and then its local segments, and repeat the process until all segments are generated. 

Although primitive, we can see that the resulting heatmap from the synthetic data is similar to the one using real data.
From the heatmap of synthetic data, 
learned indexes can also 
do well when only one dimension is hard while the other one remains easy.
In other words, learned indexes lose their edge only when both dimensions are hard and with intensive-writes,
corroborating Message~\ref{msg:hard}.

\section{Lessons Learned} 
\label{sec:lessons} 
In this section, we summarize six lessons learned from the study:

\noindent\textbf{1. 
Using ML for subspace lookup in internal layers
and using sparse-node as node design 
generally well balance performance, space, and robustness.}
This observation inspires future work to continue in this direction
to refine cost models and address limitations (e.g., write amplification).

\noindent\textbf{2. Use hard datasets judiciously.}
During the experimentation, 
we found that most real datasets are easy.
In contrast, {\tt fb} 
has been upsampled
and {\tt osm} 
is not one-dimensional in nature \cite{osm}.
Therefore, while we can use those hard datasets 
to stress test the corner cases,
index designers may not need to put too much weights on those.

\noindent\textbf{3. Concurrency control and robustness should be first-class citizens when designing a learned index, i.e., designs should be holistic. }
Otherwise, different design choices may inherently contradict each other later.
For example, LIPP’s unified node layout which mainly considered single-threaded scenarios can significantly hurt scalability
and add many branches to range scans.

\noindent\textbf{4. Future learned indexes can benefit from cache-friendly and NUMA-aware designs.}
Since we found that good learned indexes like ALEX+ can saturate memory bandwidth, reducing cache misses and NUMA-aware data placement 
are two promising optimization directions.

\noindent\textbf{5. Memory efficiency is \emph{not} a clear advantage of updatable learned indexes.}
Simply replacing traditional indexes with learned indexes may not help much in space saving. 
We recommend future learned indexes must report the end-to-end memory consumption and 
explore alternatives (e.g., compression
\cite{compress}, persistent memory \cite{apex}) to reduce DRAM pressure.

\noindent\textbf{6. Traditional indexes are not good-for-nothing.}
Traditional indexes should not be completely abandoned because of their efficiency in write-intensive workload and hard datasets, rich functionality (e.g., support for variable-length keys) and extreme robustness.

\section{So, are Updatable Learned Indexes Ready?} \label{sec:conclusion}

In this section, 
we try to answer this question from three different angles:
\emph{yesterday}, \emph{today}, and \emph{tomorrow}.

\begin{figure}
    \centering
    \includegraphics[width=0.8\linewidth]{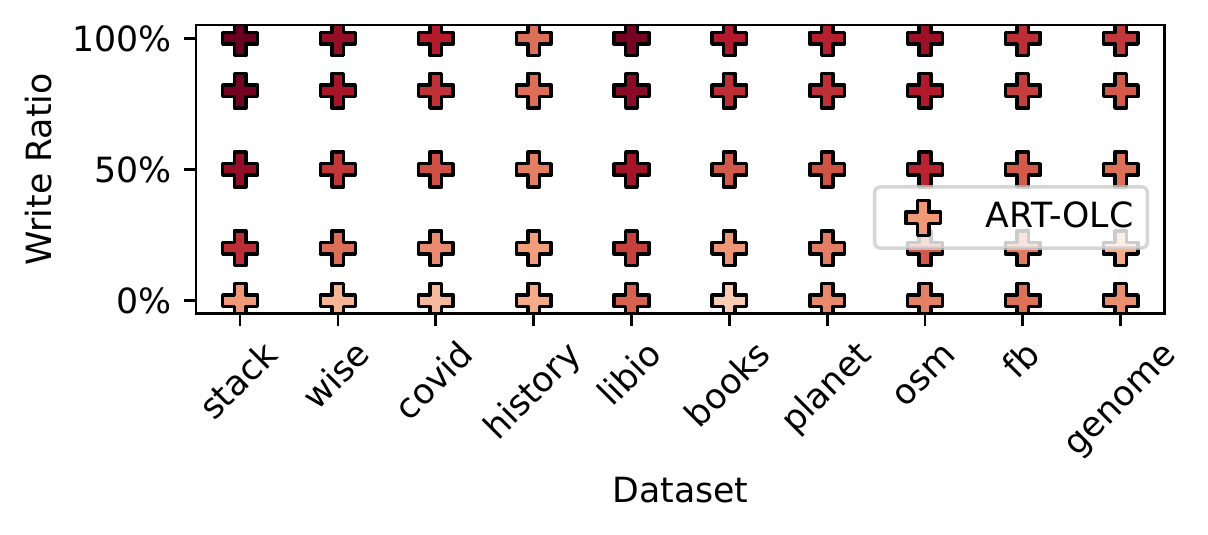}
    \caption{\added{The world without this benchmark study.  Heatmap on 24 cores with all workloads.  The datasets can't be ordered.}}
    \label{fig:before}
\end{figure}

\textbf{Yesterday.} 
If the question were asked before this benchmark study,
the answer would have been {\bf no}.
Specifically, a modern index must support concurrency, making \xindex \cite{xindex} and \finedex \cite{finedex} the only choices.
Figure~\ref{fig:before} shows what the 24-core performance heatmap would have been without our study --- ART-OLC dominates the heatmap.
Furthermore, without a quantitative metric 
on data hardness, practitioners would have no way 
to make informed decisions on when to use a learned index or not.
In addition to issues with concurrency, we observed that the supposedly fastest (single-threaded) learned index \lipp actually trades space for speed.

\textbf{Today.}  
Despite being a performance study, %
we have made contributions to parallelizing two state-of-the-art updatable learned indexes (ALEX+ and LIPP+).
We show that LIPP by design is not concurrency-friendly nor space-efficient, yet
ALEX+ is promising in terms 
of performance, scalability, space, and robustness.
Although ALEX+ has a harder time on hard datasets when the workloads are write-intensive,
most real datasets are easy and most 
real-world OLTP workloads are actually read-intensive~\cite{krueger2011fast,rehrmann2018oltpshare,lu2020performance,spanner_osdi_12}.
Hence, we conclude that ALEX+ as an updatable learned index that is {\bf almost ready}.
For write-intensive workloads, we recommend LSM-style indexing.
Some early efforts have already started to explore this direction~\cite{bourbon}.

\textbf{Tomorrow.} 
Despite the promising results, ALEX+ has limitations and  %
we recommend using it in a hardness-conscious manner.
For example,
the %
hardness of a dataset 
can be added as a new feature/dimension 
in index selection tools~\cite{selectionbench, selectionsecondary,selectionmicrosoft,selectiondb2,selectionhard}
and query optimizers~\cite{learnedoptimizer, db2optimizer, microsoftoptimizer,oracleoptimizer,Baooptimizer}.
When those components are ready,
\alexplus would also be ready.

\begin{acks}
This work is partially supported by Hong Kong General Research Fund (14200817), Hong Kong AoE/P-404/18, Innovation and Technology Fund (ITS/310/18, ITP/047/19LP) and Centre for Perceptual and Interactive Intelligence (CPII) Limited under the Innovation and Technology Fund.
\end{acks}

\clearpage
\balance
\bibliographystyle{ACM-Reference-Format}
\bibliography{cite} 

\appendix
\newpage
\renewcommand\thefigure{\Alph{figure}}
\renewcommand\thetable{\Alph{table}}  
\setcounter{figure}{0}
\section*{Appendix}

\section{Lock granularity in \alexplus} \label{sec:256lock} \RIII{R3: D1} 
\added{
We implemented two versions of ALEX+ using different lock granularities.
We finally adopted the version that uses a single optimistic lock per data node
over the one that uses one optimistic lock per 256 records in a data node.

Figure \ref{fig:alex256} shows the throughput of the two versions
under the {\tt Balanced} workload on four real datasets.
It shows that using a single optimistic lock per data node is consistently better regardless of the data hardness.  
The overhead of using a lock per 256 records is higher although it admits more concurrency.
Specifically, ALEX+ inherits ALEX's design to use exponential search within a data node and
when using one lock per 256 records, locks could be acquired in different orders as a search can go 
either direction within a data array. 
This can lead to deadlocks.
To prevent deadlock, per-256-record locking requires 
ALEX+ to release all the acquired locks once it fails to acquire a lock and restarts. 
In contrast, per-node locking is deadlock-free. 
Moreover, the lock can be inlined within a data node's header, accessing which incurs no extra cost if the header is 
smaller than a cacheline because the node header contains essential metadata (e.g., the ML model) that must be accessed anyway.}

\section{Support for Non-Unique Keys} 
\label{sec:unique}
\begin{figure}[t]
    \centering
    \includegraphics[width=\linewidth]{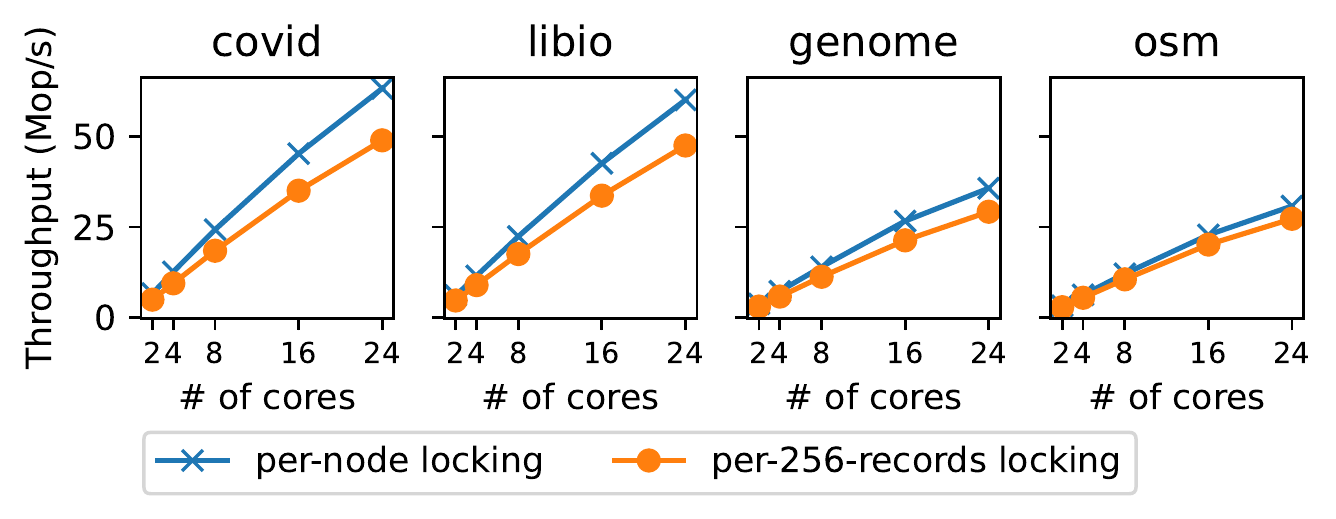}
    \caption{Throughput of \alexplus using different lock granularities under the Balanced workload.}
    \label{fig:alex256}

    \centering
    \includegraphics[width=\linewidth]{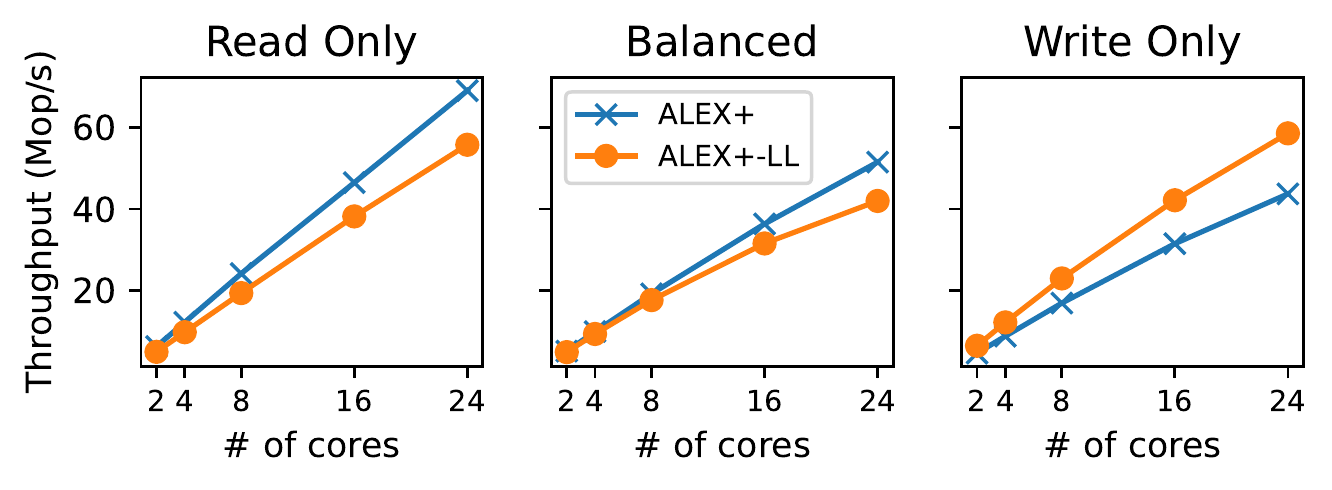}
    \caption{Throughput of \alexplus on the {\tt wiki} dataset with non-unique keys under different workloads.}
    \label{fig:alexdup}
\end{figure}

\RII{AE6 R2: W1/D3/I2}
\added{
Our evaluation has been mainly using workloads with unique keys only. 
But one may also build indexes on non-primary attributes with duplicated keys.
In this section, we explore the impact of duplicate keys on learned indexes. 

There are two main approaches for an index to handle duplicated keys: 
(1) using a linked list and (2) inlining.
For (1), values of the same key form a linked list,
and only the first occurrence of key would be presented in the nodes.
For (2), all occurrences of the values of the same key would be 
stored (inlined) in the index node.

Generally, the use of inlining or linked list is a tradeoff between lookup and insert.
Inlining favors lookup because the values
are co-located in memory, traversing which does not require additional pointer chasing.
However, inlining is less friendly to insertion and may require sophisticated space management 
(e.g., for variable-size payload)
in index nodes, increasing the write amplification as new inserts may cause the payloads and keys to be moved around in the node (e.g., to keep key-value pairs sorted). 
In contrast, the linked list approach favors insertion because 
the values are stored out-of-place; the index node only needs to store a pointer to the list of values per key.  
Yet, the linked list approach may reduce lookup throughput because of pointer chasing.

Among all the learned indexes surveyed, none but ALEX (hence ALEX+) supports duplicated keys using inlining.
Hence, to study whether learned indexes would follow the tradeoff discussed above when facing duplicated keys,
we implemented a version of ALEX+ that uses linked list to handle duplicates (denoted as ALEX+LL).  
Figure \ref{fig:alexdup} shows the performance of ALEX+ (that uses inlining) 
and ALEX+LL on SOSD's {\tt wiki} dataset, which contains duplicates.
The results show that ALEX+ as the overall best concurrent learned index also exhibits the aforementioned tradeoff,
where the use of linked list makes it better in insert but the use of inlining makes in better it lookup.
}

\section{Computing the optimal PLA-model}
\label{sec:pla}\RI{R1: W2/D2/I2}
\added{
We use the algorithm in \cite{pgm} to compute the optimal PLA-model of a given dataset.
It belongs to a family of online algorithms~\cite{pgm-geometry-paper} that aim to fit straight lines for time-series data, where each data point $(t, v)$ has a timestamp $t$ with an error range $\epsilon$ on the value $v$.
Their goal is to maintain a set of straight lines that can fit all the data points within their error ranges.
Since the data is time-series, $t_i$ is smaller than $t_j$ if $i<j$.
In \cite{pgm}, it views each value in a data array as a tuple 
$(k, r)$, where $k$ is the key and $r$ is the position in the data array.
In range indexes, it is natural that 
$k_i$ is smaller than $k_j$ if $i<j$.
Hence, the algorithm in \cite{pgm}
leverages the time-series-like total order to achieve linear time and space complexity.
Briefly, the idea is to incrementally construct a convex hull for a set of points $(k_i, r_i)$ for $i=1...n$.
The rank $r_i$ would be within $\pm\epsilon$ as long as their convex hull can be enclosed by a bounding box with height $2\epsilon$.
Incrementally updating the convex hull admits linear time and space complexity.
If a point $(k_i, r_i)$ goes beyond the bounding box, 
that means the current set of data cannot be fitted by any straight line, then the algorithm increments the number of segments by one and moves on to construct the next segment starting with $k_i$.
For more details, interested readers may refer to~\cite{pgm, pgm-geometry-paper}.}

\section{Approximating Data Hardness}
\label{sec:data}\RIII{R3: I1/I2}
\added{
All single-dimensional indexes, to our best knowledge, 
are designed as a tree of linear models.
Therefore, the PLA %
of a single-dimensional dataset that
captures the minimal number of linear models $H$ required to fit the data distribution
is a natural candidate to approximate the data hardness.
Yet, PLA has a subtle parameter $\epsilon$ that governs the approximation quality.
Intuitively, a good approximation (i.e., a good choice of $\epsilon$) 
shall align best with any learned index performance.
In other words, if a data hardness approximation metric determines that dataset $A$ is easier than dataset $B$ (i.e., $H_A < H_B$), 
then it would be a good approximation if all learned indexes 
perform better on $A$ than on $B$.
That forms our basis for choosing the $\epsilon$ values.
}

\added{
From a high-level, our goal is to choose a pair of $\epsilon$ values for local and hardness approximation, respectively. 
The best choice should lead to data hardness (both local and global) that aligns well with the index performance expectations (i.e., harder leads to lower performance). 
In general, a small/large $\epsilon$ in PLA is a fine-grained approximation of the CDF that captures the local/global non-linearity better.
Therefore, to pick a pair of suitable $\epsilon$ values, we first fix the small value, and try different large values to observe whether the resulting approximation would align with index performance behaviors. 
We repeat this process with different small values (for each of which we subsequently test different large values) to find a desirable pair eventually. 
We empirically experimented various values and found that many choices would work and provide similar results (e.g., 32/4096, 64/2048, etc.). 
For brevity, below we take 32/4096 as the small/large $\epsilon$ values (which led to the best results among our tests) to explain the rationale and process in more detail.

\begin{figure}[t]
    \centering
    \includegraphics[width=\linewidth]{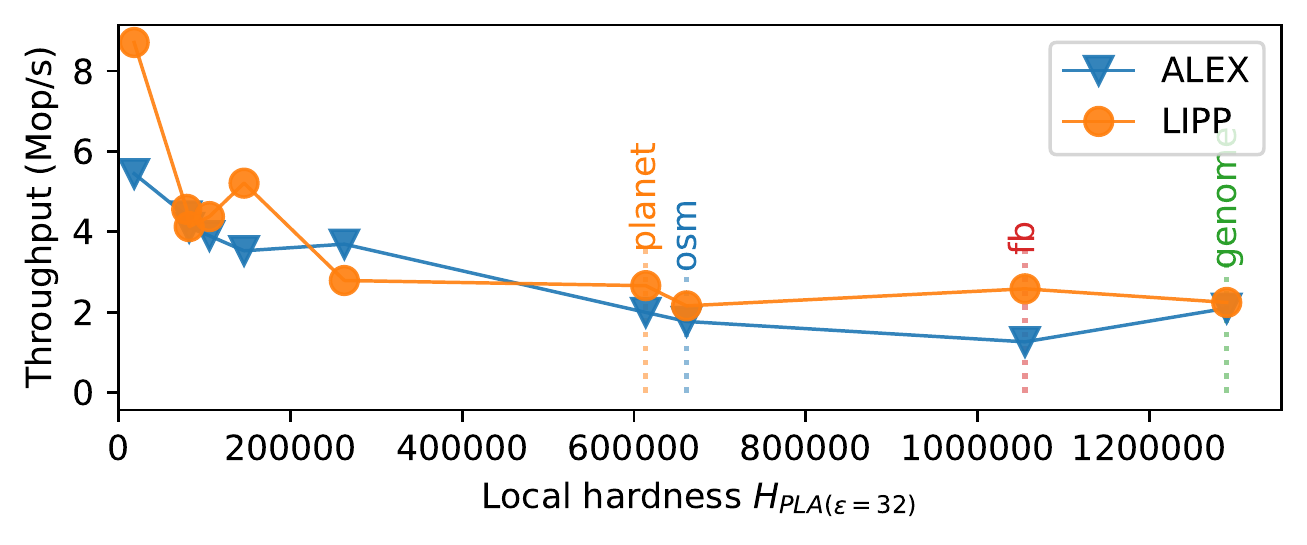}
    \caption{Throughput of the Balanced workload under varying PLA local hardness with a small $\epsilon$ value (32) .}
    \label{fig:pla32}
    \includegraphics[width=\linewidth]{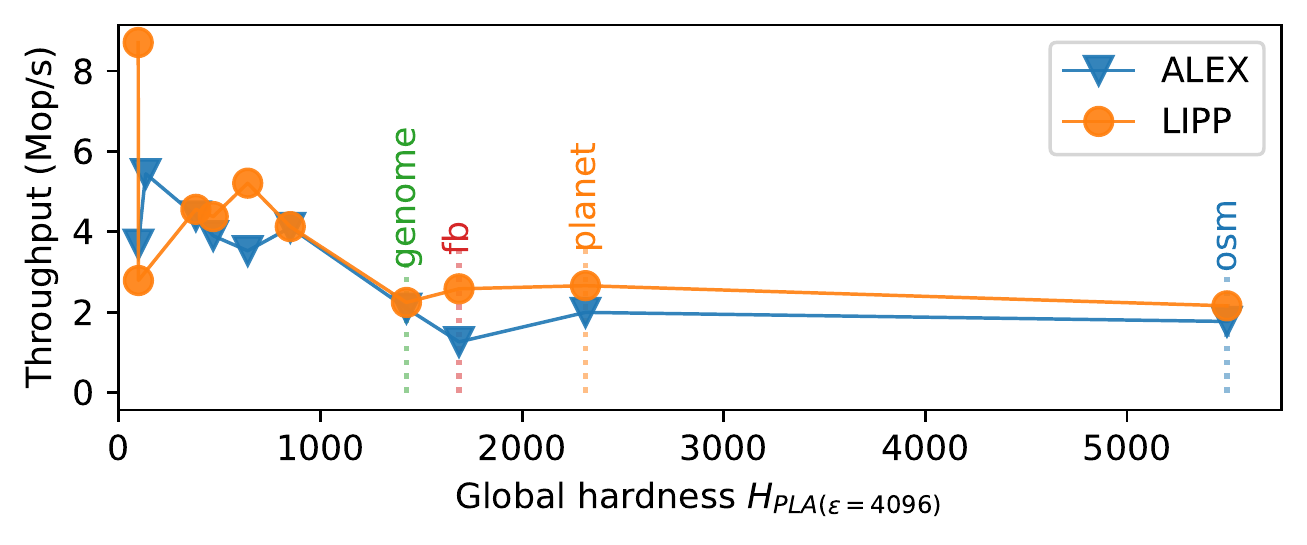}
    \caption{Throughput of the Balanced workload under vaying PLA ($\epsilon=4096$) global hardness.}
    \label{fig:pla4096}
\end{figure}

Figure~\ref{fig:pla32} shows the throughput of ALEX and LIPP 
under different data hardness based on a small $\epsilon=32$ value (other learned indexes behave similarly; omitted for clarity).
It is a fairly good approximation because 
the performance of the surveyed learned indexes align 
with the data hardness pretty well --- the index throughput generally degrades when $H$ increases.
However, we observe that the use of PLA ($\epsilon=32$) alone is insufficient 
because the indexes show similar throughput 
even though $H_{planet}$ and $H_{osm}$
are much smaller than $H_{fb}$ and $H_{genome}$ under that metric.
That essentially implies that {\tt planet} and {\tt osm} 
possess some other hardness that is not captured by PLA ($\epsilon=32$).

\begin{figure}[t]
    \centering
    \begin{subfigure}{.48\linewidth}
         \centering
    \includegraphics[width=\linewidth]{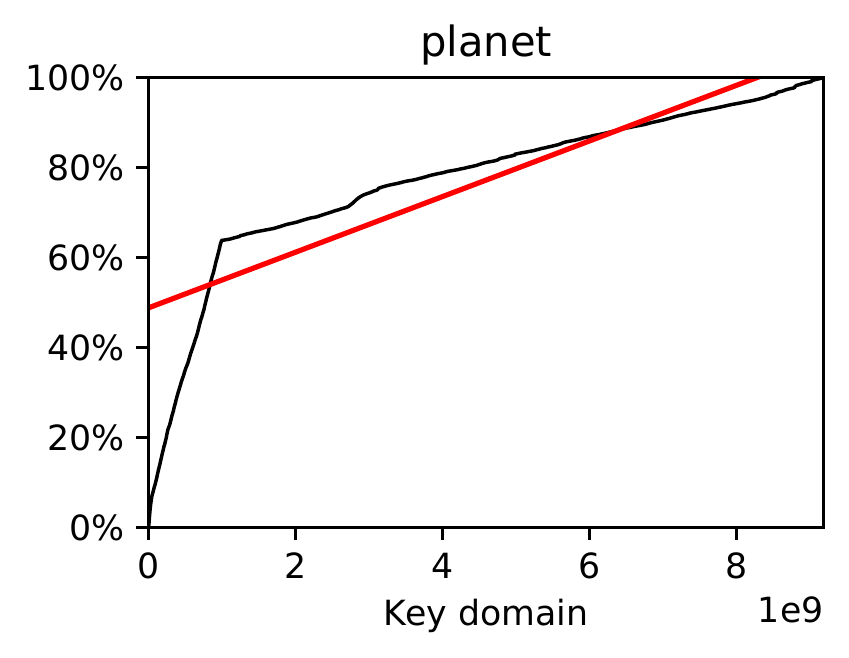}
         \caption{{\tt planet}  }
         
     \end{subfigure}
     \begin{subfigure}{.48\linewidth}
         \centering
    \includegraphics[width=\linewidth]{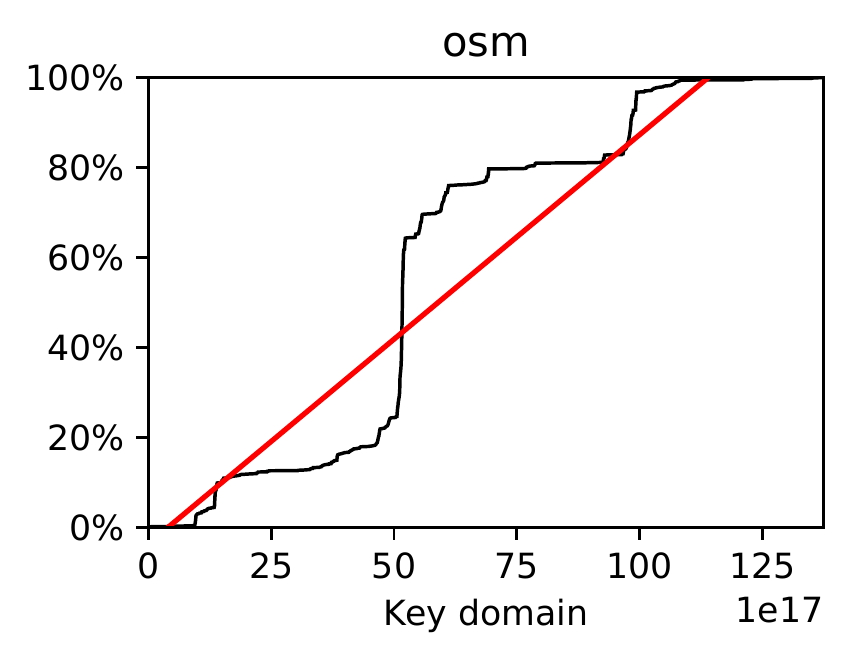}
    \caption{{\tt osm} }
     \end{subfigure}
    \caption{Fitting one linear regression model (red line) and using the MSE as the global hardness.}
    \label{fig:mse-chris}
\end{figure}

\begin{figure}[t]
\centering
    \includegraphics[width=\linewidth]{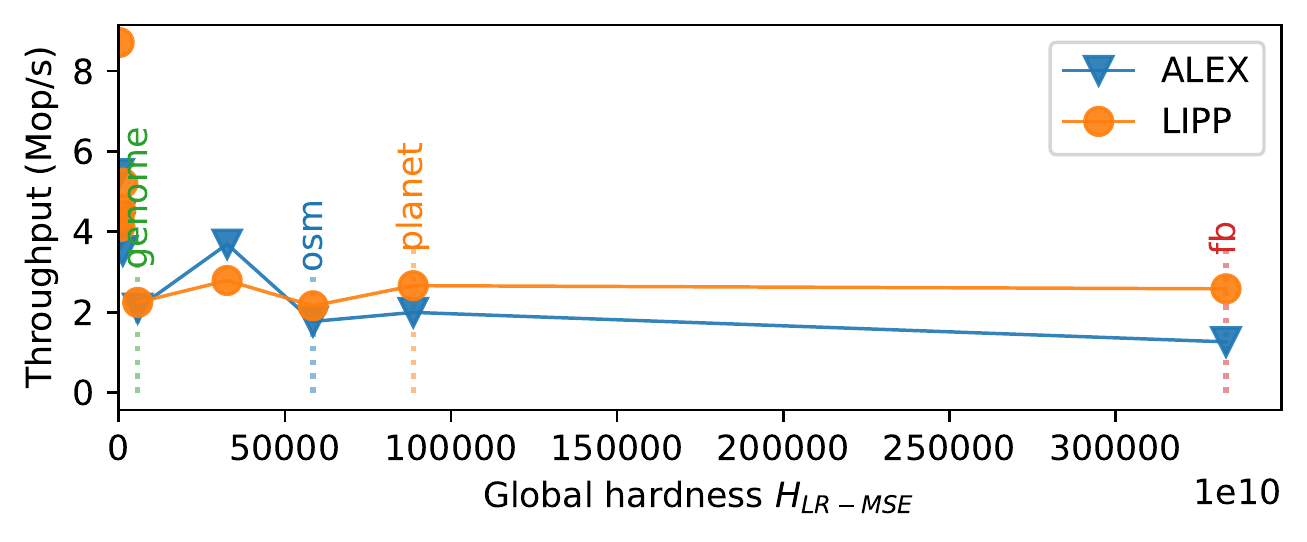}
    \caption{Throughput of the Balanced workload under varying global hardness based on MSEs.}
    \label{fig:mse}
\end{figure}

With a small $\epsilon$ capturing local hardness, now we experiment with different large values for global non-linearity. 
Figure~\ref{fig:pla4096} shows the throughput of ALEX and LIPP by arranging a dataset's global hardness using PLA with $\epsilon=4096$. 
A high $\epsilon$ value means a large step size when segmenting a CDF, which can capture a dataset's global non-linearity.
Furthermore, PLA is less sensitive to outliers (e.g., the few outliers in {\tt fb} can all be represented by a few new models, hence adding the hardness value only by a bit).
By using PLA with a large $\epsilon$, the approximation correctly ranks 
{\tt planet} and {\tt osm} as globally harder than 
{\tt fb} and {\tt genome}, complementing the local non-linearity approximation obtained using a small $\epsilon$ value.

\begin{figure*}[t]
    \centering
     \begin{subfigure}{0.48\linewidth}
         \centering
         \includegraphics[width=\linewidth]{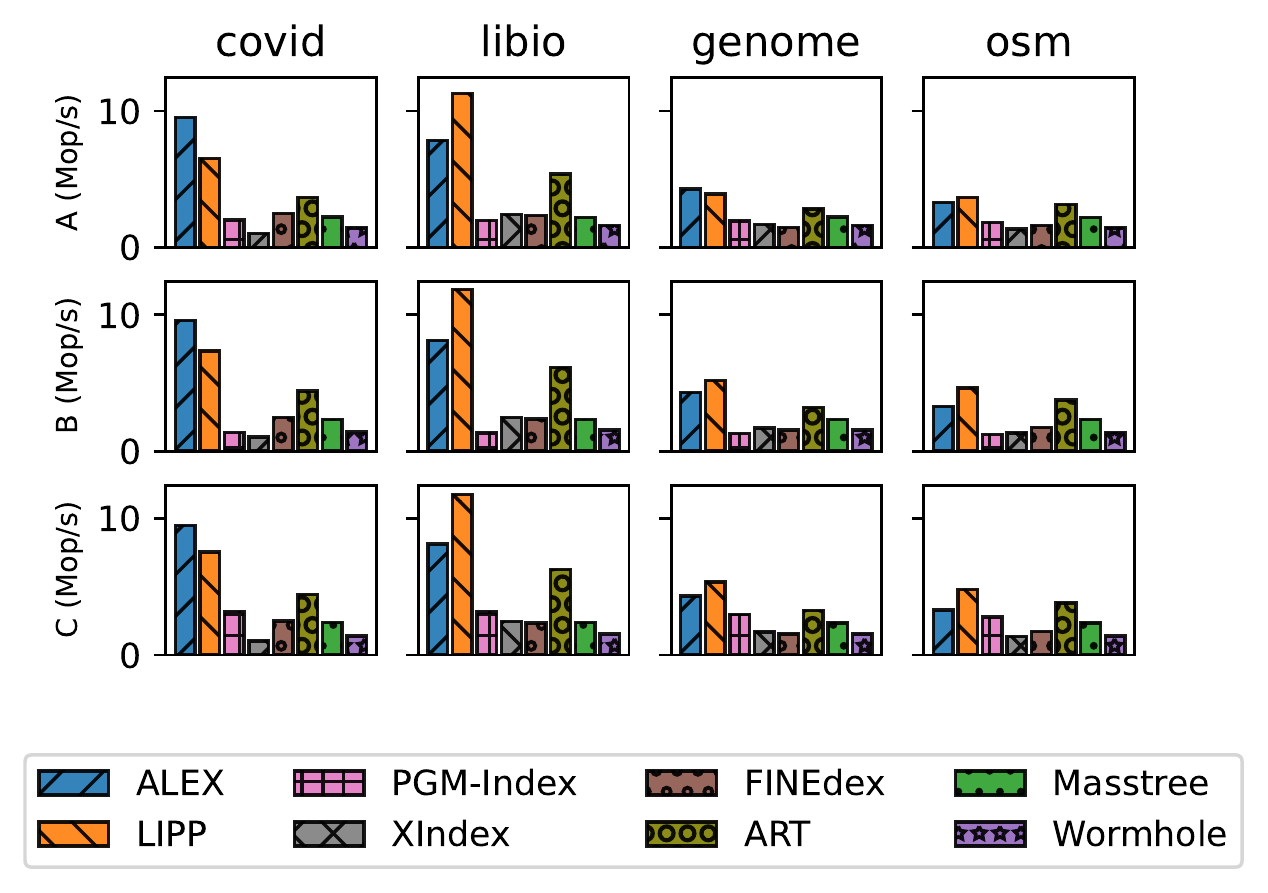}
         \caption{Single core}
     \end{subfigure}
     \begin{subfigure}{0.48\linewidth}
         \centering
         \includegraphics[width=\linewidth]{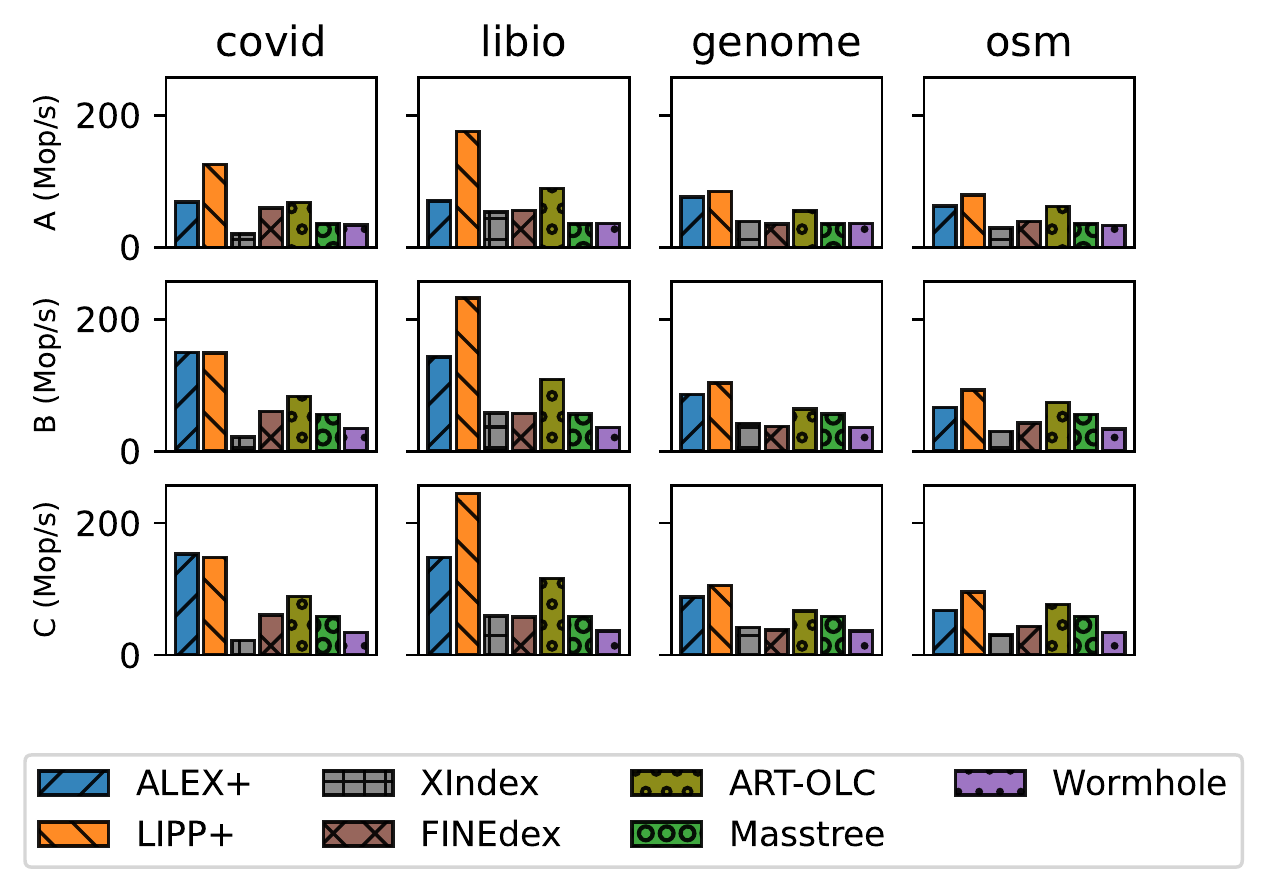}
         \caption{24 cores}
     \end{subfigure}
    \caption{Throughput of indexes under YCSB workload with Zipfian distribution.}
    \label{fig:ycsb}
\end{figure*}

One may notice that there are other alternatives to complement local non-linearity.
For example, we can fit the whole CDF using one linear regression model 
and use its mean-square-error (MSE) to measure its global non-linearity, as shown in Figure~\ref{fig:mse-chris}.
However, we note using a large $\epsilon$ value in PLA can capture the global non-linearity better than using MSEs. 
Figure~\ref{fig:mse} shows the throughput of ALEX and LIPP by arranging the datasets' global hardness using their MSEs.
In the figure, {\tt planet} and {\tt osm} are harder than {\tt genome} in global non-linearity, indicating the approach can complement the local non-linearity dimension. 
However, it still cannot explain the behavior of the indexes on {\tt fb}.
Specifically, {\tt fb} is regarded as much harder than {\tt planet} and {\tt osm} both locally (Figure \ref{fig:pla32}) and ``globally'' (Figure \ref{fig:mse}). 
One would therefore expect an index to perform worse on {\tt fb} than on {\tt osm} and {\tt planet}. 
However, we observe that for the same index (e.g., ALEX), it performs more or less the same across {\tt planet}, {\tt osm} and {\tt fb}. 
In other words, {\tt fb}'s overall hardness (considering both local and global) should be similar to the other two datasets, otherwise the index should not perform similarly across all the three datasets.
This shows the limitation of MSE: 
it is too sensitive to outliers, giving {\tt fb} (which has a few outliers with very large keys) an overly-high global non-linearity.
Yet using PLA as we explained above can model the data hardness better.  
}

\section{YCSB Results} \label{sec:ycsb}
\RIII{R3: W1}

\added{
Our evaluation has mainly focused on index performance 
under workloads whose keys are uniformly sampled 
from real data.
In this section, we study the index performance using 
workloads with non-uniform key distributions.
Specifically, we use three workloads from the industrial-strength YCSB benchmark \cite{ycsb}:
\begin{itemize}[leftmargin=*]\setlength\itemsep{0em}
\item YCSB-A: an update-heavy workload
that contains 50\% lookup requests and 50\% update requests where keys 
are chosen under a Zipfian distribution.
\item YCSB-B: similar to YCSB-A except 
it is read-heavy with 95\% of requests are lookups
and 5\% are updates.
\item YCSB-C: similar to the other two except it is a read-only workload with 100\% lookup.
\end{itemize}

The default Zipfian constant in YCSB is 0.99.
Figure \ref{fig:ycsb} shows that our findings
in the main discussion can be extended to non-uniform workloads.  
For example, ALEX and LIPP are still the leaders in most cases while ART comes close on hard data.
It is worth noting that LIPP+ remains competitive in YCSB workloads even using multiple cores
(Figure \ref{fig:ycsb}b). 
YCSB workloads have no key insertions but only updates the payloads of existing keys.
Hence, it can scale 
because there are no updates on the per-node statistics along its update path and triggering no atomic instructions.
Yet, that does not change the fact that LIPP consumes huge space, cannot scale with inserts, and is not range-scan friendly.
}

\end{document}